\documentstyle[twoside,psfig,amsmath]{article}

\catcode`\@=11
\long\def\@makefntext#1{
\protect\noindent \hbox to 3.2pt {\hskip-.9pt  
$^{{\eightrm\@thefnmark}}$\hfil}#1\hfill}		

\def\thefootnote{\fnsymbol{footnote}}
\def\@makefnmark{\hbox to 0pt{$^{\@thefnmark}$\hss}}	
	
\def\ps@myheadings{\let\@mkboth\@gobbletwo
\def\@oddhead{\hbox{}
\rightmark\hfil\eightrm\thepage}   
\def\@oddfoot{}\def\@evenhead{\eightrm\thepage\hfil
\leftmark\hbox{}}\def\@evenfoot{}
\def\sectionmark##1{}\def\subsectionmark##1{}}

\oddsidemargin=\evensidemargin
\renewcommand{\thefootnote}{\fnsymbol{footnote}}

\newcounter{sectionc}\newcounter{subsectionc}\newcounter{subsubsectionc}
\renewcommand{\section}[1] {\vspace{12pt}\addtocounter{sectionc}{1} 
\setcounter{subsectionc}{0}\setcounter{subsubsectionc}{0}\noindent 
	{\tenbf\thesectionc. #1}\par\vspace{5pt}}
\renewcommand{\subsection}[1] {\vspace{12pt}\addtocounter{subsectionc}{1} 
	\setcounter{subsubsectionc}{0}\noindent 
	{\bf\thesectionc.\thesubsectionc. {\kern1pt \bfit #1}}\par\vspace{5pt}}
\renewcommand{\subsubsection}[1] {\vspace{12pt}\addtocounter{subsubsectionc}{1}
	\noindent{\tenrm\thesectionc.\thesubsectionc.\thesubsubsectionc.
	{\kern1pt \tenit #1}}\par\vspace{5pt}}
\newcommand{\nonumsection}[1] {\vspace{12pt}\noindent{\tenbf #1}
	\par\vspace{5pt}}

\newcounter{appendixc}
\newcounter{subappendixc}[appendixc]
\newcounter{subsubappendixc}[subappendixc]
\renewcommand{\thesubappendixc}{\Alph{appendixc}.\arabic{subappendixc}}
\renewcommand{\thesubsubappendixc}
	{\Alph{appendixc}.\arabic{subappendixc}.\arabic{subsubappendixc}}

\renewcommand{\appendix}[1] {\vspace{12pt}
        \refstepcounter{appendixc}
        \setcounter{figure}{0}
        \setcounter{table}{0}
        \setcounter{lemma}{0}
        \setcounter{theorem}{0}
        \setcounter{corollary}{0}
        \setcounter{definition}{0}
        \setcounter{equation}{0}
        \renewcommand{\thefigure}{\Alph{appendixc}.\arabic{figure}}
        \renewcommand{\thetable}{\Alph{appendixc}.\arabic{table}}
        \renewcommand{\theappendixc}{\Alph{appendixc}}
        \renewcommand{\thelemma}{\Alph{appendixc}.\arabic{lemma}}
        \renewcommand{\thetheorem}{\Alph{appendixc}.\arabic{theorem}}
        \renewcommand{\thedefinition}{\Alph{appendixc}.\arabic{definition}}
        \renewcommand{\thecorollary}{\Alph{appendixc}.\arabic{corollary}}
        \renewcommand{\theequation}{\Alph{appendixc}.\arabic{equation}}
        \noindent{\tenbf Appendix \theappendixc #1}\par\vspace{5pt}}
\newcommand{\subappendix}[1] {\vspace{12pt}
        \refstepcounter{subappendixc}
        \noindent{\bf Appendix \thesubappendixc. {\kern1pt \bfit #1}}
	\par\vspace{5pt}}
\newcommand{\subsubappendix}[1] {\vspace{12pt}
        \refstepcounter{subsubappendixc}
        \noindent{\rm Appendix \thesubsubappendixc. {\kern1pt \tenit #1}}
	\par\vspace{5pt}}

\newcommand{\textlineskip}{\baselineskip=13pt}
\newcommand{\smalllineskip}{\baselineskip=10pt}

\def\eightcirc{
\begin{picture}(0,0)
\put(4.4,1.8){\circle{6.5}}
\end{picture}}
\def\eightcopyright{\eightcirc\kern2.7pt\hbox{\eightrm c}} 

\newcommand{\copyrightheading}[1]
	{\vspace*{-2.5cm}\smalllineskip{\flushleft
	{\footnotesize International Journal of Modern Physics B, #1}\\
	{\footnotesize $\eightcopyright$\, World Scientific Publishing
	 Company}\\
	 }}

\newcommand{\publisher}[2]{{\begin{center}\footnotesize\smalllineskip 
	Received #1\\
	Revised #2
	\end{center}
	}}

\def\abstracts#1#2#3{{
	\centering{\begin{minipage}{4.5in}\baselineskip=10pt\footnotesize
	\parindent=0pt #1\par 
	\parindent=15pt #2\par
	\parindent=15pt #3
	\end{minipage}}\par}} 

\def\keywords#1{{
	\centering{\begin{minipage}{4.5in}\baselineskip=10pt\footnotesize
	{\footnotesize\it Keywords}\/: #1
	\end{minipage}}\par}}

\renewenvironment{thebibliography}[1]			
	{\frenchspacing
	 \ninerm\baselineskip=11pt
	 \begin{list}{\arabic{enumi}.}
	{\usecounter{enumi}\setlength{\parsep}{0pt}
	 \setlength{\leftmargin 12.7pt}{\rightmargin 0pt} 
	 \setlength{\itemsep}{0pt} \settowidth
	{\labelwidth}{#1.}\sloppy}}{\end{list}}

\newcounter{itemlistc}
\newcounter{romanlistc}
\newcounter{alphlistc}
\newcounter{arabiclistc}

\newcommand{\fcaption}[1]{
        \refstepcounter{figure}
        \setbox\@tempboxa = \hbox{\footnotesize Fig.~\thefigure. #1}
        \ifdim \wd\@tempboxa > 5in
           {\begin{center}
        \parbox{5in}{\footnotesize\smalllineskip Fig.~\thefigure. #1}
            \end{center}}
        \else
             {\begin{center}
             {\footnotesize Fig.~\thefigure. #1}
              \end{center}}
        \fi}

\newcommand{\tcaption}[1]{
        \refstepcounter{table}
        \setbox\@tempboxa = \hbox{\footnotesize Table~\thetable. #1}
        \ifdim \wd\@tempboxa > 5in
           {\begin{center}
        \parbox{5in}{\footnotesize\smalllineskip Table~\thetable. #1}
            \end{center}}
        \else
             {\begin{center}
             {\footnotesize Table~\thetable. #1}
              \end{center}}
        \fi}

\def\@citex[#1]#2{\if@filesw\immediate\write\@auxout
	{\string\citation{#2}}\fi
\def\@citea{}\@cite{\@for\@citeb:=#2\do
	{\@citea\def\@citea{,}\@ifundefined
	{b@\@citeb}{{\bf ?}\@warning
	{Citation `\@citeb' on page \thepage \space undefined}}
	{\csname b@\@citeb\endcsname}}}{#1}}

\newif\if@cghi
\def\cite{\@cghitrue\@ifnextchar [{\@tempswatrue
	\@citex}{\@tempswafalse\@citex[]}}
\def\citelow{\@cghifalse\@ifnextchar [{\@tempswatrue
	\@citex}{\@tempswafalse\@citex[]}}
\def\@cite#1#2{{$\null^{#1}$\if@tempswa\typeout
	{IJCGA warning: optional citation argument 
	ignored: `#2'} \fi}}

\def\pmb#1{\setbox0=\hbox{#1}
	\kern-.025em\copy0\kern-\wd0
	\kern.05em\copy0\kern-\wd0
	\kern-.025em\raise.0433em\box0}

\def\fnt#1#2{\footnotetext{\kern-.3em
	{$^{\mbox{\scriptsize #1}}$}{#2}}}

\def\fpage#1{\begingroup
\voffset=.3in
\thispagestyle{empty}\begin{table}[b]\centerline{\footnotesize #1}
	\end{table}\endgroup}

\def\runninghead#1#2{\pagestyle{myheadings}
\markboth{{\protect\footnotesize\it{\quad #1}}\hfill}
{\hfill{\protect\footnotesize\it{#2\quad}}}}
\font\tenrm=cmr10
\font\tenit=cmti10 
\font\tenbf=cmbx10
\font\bfit=cmbxti10 at 10pt
\font\ninerm=cmr9

\font\eightrm=cmr8

\def\qed{\hbox{${\vcenter{\vbox{			
   \hrule height 0.4pt\hbox{\vrule width 0.4pt height 6pt
   \kern5pt\vrule width 0.4pt}\hrule height 0.4pt}}}$}}

\renewcommand{\thefootnote}{\fnsymbol{footnote}}	

\def\bsc{{\sc a\kern-6.4pt\sc a\kern-6.4pt\sc a}}	
\def\bflatex{\bf L\kern-.30em\raise.3ex\hbox{\bsc}\kern-.14em 
T\kern-.1667em\lower.7ex\hbox{E}\kern-.125em X} 

\newcommand{\prl}{Phys. Rev. Lett. }
\newcommand{\prb}{Phys. Rev. B }

\newcommand{\pr}{Phys. Rev. }
\newcommand{\jpsj}{J. Phys. Soc. Jpn. }
\newcommand{\jpa}{J. Phys. A }
\newcommand{\calhat}[1]{\hat{\cal {#1}}}

\begin{document}

\runninghead {Quantum Effects and Broken Symmetries in Frustrated Antiferromagnets} 
{Quantum Effects and Broken Symmetries in Frustrated Antiferromagnets}

\normalsize\textlineskip
\thispagestyle{empty}
\setcounter{page}{1}

\copyrightheading{}			

\vspace*{0.88truein}

\fpage{1}
\centerline{\bf Quantum Effects and Broken Symmetries }
\vspace*{0.035truein}
\centerline{\bf in Frustrated Antiferromagnets}
\vspace*{0.37truein}

\centerline{\footnotesize LUCA CAPRIOTTI}
 
\vspace*{0.015truein}
\centerline{\footnotesize\it Istituto Nazionale per la Fisica della Materia
(INFM), Unit\`a di Firenze, } 
\baselineskip=10pt
\centerline{\footnotesize\it Largo E. Fermi 2, I-50125 Firenze, Italy}
\vspace*{10pt}
\vspace*{0.225truein}
\publisher{(received date)}{(revised date)}

\vspace*{0.21truein}

\abstracts{We investigate the interplay between frustration and zero-point quantum
fluctuations in the ground state of the triangular and $J_1{-}J_2$ Heisenberg
antiferromagnets, using finite-size spin-wave theory, exact diagonalization,
and quantum Monte Carlo methods.
In the triangular Heisenberg antiferromagnet, by performing
a systematic size-scaling analysis, we have obtained strong evidences
for a gapless spectrum and a finite value of the thermodynamic
order parameter, thus confirming the existence of long-range
N\'eel order.
The good agreement between the finite-size spin-wave
results and the exact and quantum Monte Carlo data also supports the
reliability of the spin-wave expansion to describe both
the ground state and the low-energy spin excitations of the
triangular Heisenberg antiferromagnet.
In the $J_1{-}J_2$ Heisenberg model, our results indicate
the opening of a finite gap in the thermodynamic excitation spectrum at
$J_2/J_1 \simeq 0.4$, marking the melting of
the antiferromagnetic N\'eel order and the onset of a non-magnetic ground
state.
In order to characterize the nature of the latter quantum-disordered phase
we have computed the susceptibilities
for the most important crystal symmetry breaking operators.
In the ordered phase the effectiveness of the spin-wave theory in
reproducing the low-energy excitation spectrum suggests that the uniform spin
susceptibility
of the model is very close to the linear spin-wave prediction. }{}{}

\vspace*{10pt}


\vspace*{1pt}\textlineskip	
\section{Introduction}	
\vspace*{-0.5pt}
\noindent

\textheight=7.8truein
\setcounter{footnote}{0}
\renewcommand{\thefootnote}{\alph{footnote}}

The physics of quantum antiferromagnets is a very old topic, 
dating back to the early days of quantum mechanics itself. 
Nonetheless, after many years of intensive study, the interest in 
this research field is still high, with several new problems
arising from the behavior of low-dimensional magnetic materials.
This is also due to the existence of simple toy-models in which the 
interplay between antiferromagnetism, symmetry, dimensionality
and strong quantum correlations leads to fascinating effects in the
low-temperature physics, often reproducing the behavior of real systems.
Among them, the nearest-neighbor Heisenberg Hamiltonian, 
\begin{equation}
\calhat{H}=J \sum_{n.n.} \,
\hat{{\bf {S}}}_{i} \cdot \hat{{\bf {S}}}_{j}~,
\label{eq.heisham}
\end{equation}
where ${\bf \hat{S}}_{i}=(\hat{S}^x_i,\hat{S}^y_i,\hat{S}^z_i)$
are spin-$s$ operators and $J$ is the (positive) exchange integral, has certainly played
a central role as an ideal test ground to investigate the influence of quantum 
effects on the mechanism of spontaneous symmetry breaking. 
In fact, in contrast to the ferromagnetic state, the classical N\'eel state
is not an eigenstate of the Heisenberg Hamiltonian and therefore, in general,
the ground state of the latter does not have a purely classical representation.
Hence, quantum effects may play an important role in modifying the zero-temperature
properties of the model from the classical ($s\to\infty$) limit.
In particular, reduced dimensionality and a small spin value might enhance 
zero-point quantum fluctuations up to the point of destroying the classical N\'eel order,
thus stabilizing a ground state with symmetries and correlations different
from its classical counterpart.

Indeed, in one dimension and for $s=1/2$, a famous exact solution found by Bethe in 1931 (Ref.~\cite{bethe}) 
showed that quantum effects prevent the onset of true long-range 
antiferromagnetic order, giving instead a power-law decay of the 
spin-spin correlation functions. Despite Bethe's promise to generalize
his solution to the two-dimensional square lattice case, appearing in the conclusions
of his paper, this was never done, and the issue of the existence of long-range order 
in the ground state of the two-dimensional Heisenberg model has been left 
unsolved for many years.
The rigorous proof of the ordered nature of the ground state of the square 
Heisenberg antiferromagnet was given in fact,  for $s\ge 1$, only in 1986,\cite{neves}
and has not been extended yet to the spin-half case where zero-point quantum fluctuations are stronger.

This problem  became a {\em hot topic} when possible connections between a non-magnetic ground state and
the mechanism of high-$T_c$ superconductivity were put forward by Anderson in 1987.\cite{anderson2}
In fact, since the stoichiometric compounds of the high-$T_c$ superconductors
are good realizations of a $s=1/2$ square Heisenberg antiferromagnet,
this conjecture focused the attention on the properties of this system.
Fortunately enough, at that time the development of modern computers was such that
the use of numerical techniques could compensate for the lack of exact analytical results.
In particular, quantum Monte Carlo methods have been of crucial importance,
by allowing one to perform a systematic size-scaling of the physical observables 
and therefore to reach a definite conclusion.\cite{reger} 
As a result, even if a rigorous proof is still lacking,
there is at present a general consensus about the ordered nature
of the ground state of the spin-half square Heisenberg antiferromagnet:
in two dimensions, reduced dimensionality and a low spin value
do not seem enough to stabilize, within the Heisenberg model, a non-magnetic ground state.

Better candidates for a realization of disordered ground states in two dimensions are
frustrated spin models. In these systems, in fact, the usual antiferromagnetic alignment 
between spins is hindered by the geometry of the lattice or by the presence 
of competing interactions. 
As a result, a general feature introduced by frustration is a less stable classical minimum energy 
configuration which is more likely to be destabilized by zero-point quantum fluctuations 
for a small spin value.
Among this class of systems two prototypical
examples are given by the triangular Heisenberg antiferromagnet, and
the $J_1{-}J_2$ Heisenberg model. The nature of the ground state in these
frustrated spin models represents the main topic of this paper.

\begin{figure}
\centerline{\psfig{bbllx=180pt,bblly=310pt,bburx=440pt,bbury=485pt,%
figure=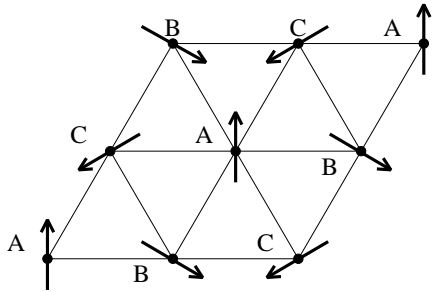,width=60mm,angle=0}}
\caption{The classical N\'eel state consists of coplanar spins
forming $\pm 2\pi/3$ angles between nearest neighbors.
This leads to a  $\protect\sqrt{3}\times\protect\sqrt{3}$ periodicity with
the spins on the three sublattices A,B,C ferromagnetically aligned.  }
\label{fig.neeltaf}
\end{figure}

The triangular Heisenberg antiferromagnet is described by the Hamiltonian (\ref{eq.heisham}), 
where $i$ and $j$ are the sites of a triangular lattice. 
Due to the geometry of the lattice (see Fig.~\ref{fig.neeltaf}), the classical 
minimum energy configuration of this model is not the usual N\'eel state 
with antiparallel spins on neighboring sites. 
In fact, if two spins on an elementary triangular plaquette minimize their exchange 
energy by aligning antiparallel, the third one cannot do the same 
because it cannot be antiparallel to both of them, simultaneously. 
As a result, the minimum energy configuration consists of coplanar spins forming
$\pm 2\pi/3$ angles between nearest-neighbors and 
this leads to a $\sqrt{3}\times\sqrt{3}$ periodic N\'eel state 
with the spins ferromagnetically aligned on each of the three sublattices
(Fig.~\ref{fig.neeltaf}).
The resulting state, having an energy per bond twice than the optimal one,
is far less stable than that on the square lattice.

\begin{figure}
\centerline{\psfig{bbllx=170pt,bblly=310pt,bburx=450pt,bbury=490pt,%
figure=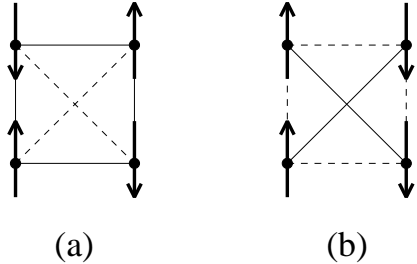,width=70mm,angle=0}}
\caption{\baselineskip .185in
The two sublattices N\'eel (a) and the collinear (b) classical states.}
\label{fig.neelj1j2}
\end{figure}

In the $J_{1}{-}J_{2}$ model, instead, frustration arises on the square lattice 
because of the presence of competing interactions, the Hamiltonian being
\begin{equation}
\calhat{H}=J{_1}\sum_{n.n.}
\hat{{\bf {S}}}_{i} \cdot \hat{{\bf {S}}}_{j}
+ J{_2}\sum_{n.n.n.}
\hat{{\bf {S}}}_{i} \cdot \hat{{\bf {S}}}_{j}~~,
\label{eq.j1j2ham}
\end{equation}
where $J_{1}$ and $J_{2}$ are the antiferromagnetic couplings between 
nearest- and next-nearest-neighbors, respectively.
Classically, the minimum energy configuration has the conventional
N\'eel order for $J_2/J_1<0.5$ [Fig.~\ref{fig.neelj1j2}~(a)].
By increasing further the frustrating interaction $J_2$ this configuration is destabilized and,
for $J_2/J_1>0.5$, the system decouples into two N\'eel ordered sublattices.
At the purely classical level, the energy of the latter configuration is independent of
the relative orientations of the staggered magnetizations on the two sublattices.
However, this degeneracy is partially lifted
by zero-point quantum fluctuations even at the lowest order in $1/s$ so that
in the $s\to \infty$ limit the minimum energy configuration is the so-called {\em collinear}
state [Fig.~\ref{fig.neelj1j2}~(b)]
with the spin ferromagnetically aligned in one direction and antiferromagnetically
in the other, corresponding to a magnetic wavevector 
${\bf Q}=(\pi,0)$ or ${\bf Q}=(0,\pi)$.\cite{chandra2}
Exactly at $J_2/J_1=0.5$ any classical state having zero total spin on each elementary square
plaquette is a minimum of the total energy. These states include both the N\'eel
and the collinear states but also many others with no long-range order. The occurrence
of a non-magnetic ground state in the quantum case,  for a small spin value,
is therefore likely around this value of the $J_2/J_1$ ratio.

The recent experimental finding of real compounds described
by the triangular and the $J_{1}{-}J_{2}$ Heisenberg antiferromagnets have 
renewed the interest in these frustrated spin systems.
In particular, the K/Si(111)-$\sqrt{3}\times\sqrt{3}$-B interface\cite{weitering,hellberg}
has turned out to be a good experimental realization of a spin-half Heisenberg antiferromagnet 
on a triangular lattice. In fact, due to strong electronic correlations, the surfaces states 
consist of a triangular arrangement of half-filled dangling bonds, which are localized and carry
local $s=1/2$ magnetic moments coupled antiferromagnetically.
Recent experimental realizations of the spin-half $J_1{-}J_2$ Heisenberg model have been
found instead in the ${\rm Li}_2{\rm VOSiO}_4$ and ${\rm Li}_2{\rm VOGeO}_4$ compounds.\cite{melzi}
These are three-dimensional systems formed by stacked square planes of $V^{4+}$ ($s=1/2$) ions
with a weak inter-plane interaction.
The structure of the $V^{4+}$ planes suggests that both the superexchange
couplings between first and second neighbors can be significant and indeed, the first
experimental results have indicated that these antiferromagnetic couplings are of the same order
of magnitude.\cite{melzi} In addition, the possibility of performing measurements under pressure
will also allow one in the near future\cite{carretta} to tune the $J_2/J_1$ ratio
and to investigate the properties of these systems in various regimes of frustration.

In this work the problem of the nature of the ground state of these frustrated spin systems
is tackled using various techniques, namely:
the finite-size spin-wave theory, exact diagonalization of small clusters by the Lanczos algorithm,
and several zero-temperature quantum Monte Carlo  methods.
The finite-size spin-wave theory has been recently proposed\cite{zhong} 
as a generalization of one of the oldest analytical techniques in the study of 
quantum magnetism.\cite{anderson3} In particular, this spin-wave expansion allows one 
to deal with finite clusters while avoiding the spurious Goldstone modes divergences 
in a straightforward way. 
Even if these results are biased by the long-range order hypothesis, 
nevertheless useful information on the thermodynamic ground state can be extracted from
the comparison with the numerical results on finite systems or
from the occurrence of a breakdown of the $1/s$ expansion.

The Lanczos method\cite{dagotto} allows the exact evaluation 
of static and dynamical properties of the finite-size system 
and, especially when combined with a careful analysis
of the symmetry of the low-energy excited states,\cite{bernu} 
can provide clear indications about the nature of the ground state.
However, due to memory constraints, exact diagonalizations techniques are limited 
in two dimensions to very small clusters ($\sim 30$ sites) 
so that it is in general difficult to perform a systematic
size scaling of the important physical observables. 
In order to numerically investigate larger systems, 
different approaches are therefore necessary.

In the unfrustrated cases, quantum Monte Carlo has turned out 
to be an essential instrument for studying  both the ground-state and 
the finite-temperature properties of a quantum antiferromagnet.\cite{manousakis}
Unfortunately, in the frustrated cases 
standard stochastic techniques cannot be applied, as their reliability is strongly limited by the
well-known {\em sign problem}. This numerical instability 
originates from the vanishing of the signal-to-noise ratio 
in the Monte Carlo sampling which occurs within bosonic models 
in the presence of frustration or, in general, in fermionic systems.

Presently, the sign problem can be controlled only at the price of introducing
some kind of approximation.  Apart from purely variational calculations,  
the simplest approximation scheme in the framework of one of the 
most efficient zero-temperature algorithms
-- the Green function Monte Carlo\cite{trivedi} -- is the fixed-node (FN) technique.\cite{fn} 
In this technique, the exact imaginary time propagator $e^{-\tau \calhat{H}}$ -- used
to filter out the ground state from the best variational guess $|\psi_G\rangle$ --
is replaced by an approximate propagator $e^{-\tau \calhat{H}_{\rm FN}}$
such that the nodes of the propagated
state $e^{-\tau \calhat{H}_{\rm FN} } |\psi_G\rangle  $ do not change,
due to  an  appropriate
choice of the effective FN Hamiltonian $\calhat{H}_{\rm FN}$
(which in turn depends on $|\psi_G\rangle$).
The FN approximation becomes exact if the so called {\em guiding} wavefunction
$|\psi_G\rangle$ is the exact ground state.
However, the fixed-node results are usually strongly biased by this ansatz
so that it is in general difficult to extract reliable
information about the ground-state correlations whenever they are not well reproduced
by the variational guess.

In order to overcome this difficulty, here we have used 
the recently developed Green function Monte Carlo with Stochastic Reconfiguration 
(GFMCSR),\cite{ss,sr} 
which allows one to release the fixed-node approximation in a controlled way 
and to obtain much more accurate estimates of the ground-state
correlations,  thus reproducing also ground-state properties that are not 
contained at the variational level in the guiding wavefunction. 
During each short imaginary time evolution  $\tau \to  \tau + \Delta \tau$,
where both the exact and the approximate propagation can be performed without sign problem
instabilities, the FN dynamic is systematically improved by requiring
that a  given  number $p$ of {\em mixed averages}\cite{ss,sr} of correlation
functions are  propagated consistently  with the exact dynamic.
By increasing the number of correlation functions one typically
improves the accuracy of the calculation  since the method becomes  exact
if all the independent correlation functions are included in the
stochastic reconfiguration (SR) scheme.
Typically,\cite{triang,j1j2}
few correlation functions ($p \sim 10$) allow to obtain
rather accurate values of the ground-state energy
with an error much less than 1$\%$ -- for lattice sizes ($N\leq 36$)
where the exact solution is available numerically --
and without a sizable loss of accuracy with increasing size.
Such accuracy is usually enough to reproduce some physical features that
are not contained at the variational level. 

In this paper, using these numerical techniques,
after a brief review of the basic concepts of spontaneous symmetry breaking 
in a quantum antiferromagnet (Sec.~2),
we provide two clear examples of systems in which 
the combined effect of frustration and quantum fluctuations 
{\em do not} or {\em do} change the zero-temperature long-range properties 
of their classical counterparts. In particular, we find that
the thermodynamic ground state of the spin-half triangular Heisenberg 
antiferromagnet (Sec.~3) is most likely long-range ordered although with a remarkable reduction of the order 
parameter with respect to the classical case. On the other hand, in the $J_{1}-J_{2}$ 
Heisenberg model (Sec.~4), 
quantum fluctuations turn out to be strong enough to melt the antiferromagnetic 
N\'eel order,  driving the ground state into a non-magnetic phase of purely 
quantum-mechanical nature.

\section{Spontaneous Symmetry Breaking in a Quantum Antiferromagnet \label{chap.ssb}}

In this section we will review the basic concepts concerning the mechanism of the
spontaneous symmetry breaking in a quantum antiferromagnet that will be
used in the present work to investigate the ground-state properties 
of the triangular and of the $J_1{-}J_2$ Heisenberg models.
Starting from the Lieb-Mattis theorem for the bipartite
Heisenberg antiferromagnet, we will introduce some general features
of the finite-size spectrum of the Heisenberg Hamiltonian.
In particular, we will focus on the importance of the structure of the low-lying excited states,
explaining also how the finite-size ground-state properties can be consistent 
with a broken symmetry in the thermodynamic limit.

\subsection{The Lieb-Mattis property}
\label{sec.lmthm}

One of the few rigorous results on the ground-state properties of the Heisenberg 
model on a bipartite lattice is the Lieb-Mattis theorem. 
Here we will reproduce the demonstration of this important result, following
the paper by E. Lieb and D. Mattis\cite{lieb} who extended and generalized
the original results obtained by W. Marshall.\cite{marshall} 

Let us consider the Heisenberg Hamiltonian
\begin{equation}
\calhat{H}= \sum_{(i,j)} J_{ij}~\hat{{\bf {S}}}_{i} \cdot \hat{{\bf {S}}}_{j}~,
\label{eq.heis}
\end{equation}
where the sum runs over all the bonds on a $d$-dimensional {\em bipartite} lattice,
$\hat{{\bf {S}}}_{i}$ are spin-$s$ operators; $J_{ij}$ is the (symmetric) 
exchange matrix such that $J_{ij}>0$, if $i$ and $j$ belong to 
different sublattices, and $J_{ij}<0$, otherwise.
We will assume that the Hamiltonian cannot
be split into sets of noninteracting spins, restricting also
for simplicity, to the case in which the number of sites 
of the two sublattices is the same. In the following $N$ will denote
the total number  of sites of the lattice.

Since the Hamiltonian (\ref{eq.heis}) commutes with all the
three components of the total spin operator, 
\begin{equation}
\hat{{\bf S}} =\sum_{i} \hat{{\bf S}}_i~, 
\end{equation}
it is known from the theory of the angular momentum that
we can construct two operators which commute with each other
and with $\calhat{H}$. For example, choosing the quantization
axis along the $z$-direction, we can consider the total spin squared, $\hat{\bf S}^2$, 
and its component along the $z$-axis, $\hat{S}^z$, 
whose eigenvalues $S(S+1)$ and $M$ are good quantum
numbers for the eigenstates of the Hamiltonian, 
\begin{equation}
|\psi_n \rangle = |n, S, M, \cdots\rangle~,
\end{equation}
such that $\calhat{H}|\psi_n \rangle = E_n |\psi_n \rangle$.
If the couplings $J_{ij}$ are translationally or rotationally invariant, also the
lattice momentum and the eigenvalues of the generators of the crystal point group
label the eigenstates of the Hamiltonian. 
However this restriction is not needed to derive the following results.

\subsubsection{Marshall-Peierls sign rule}
\label{sec.msc}

In the hypothesis stated above, the first strong result one can prove is about
the signs of the coefficients of the expansion of the ground state
of (\ref{eq.heis}) in the so-called Ising basis
whose states are specified by assigning the value of the $S^z_i$ at each lattice 
site, i.e., $|x\rangle=\prod_{i=1} |m_i\rangle$ with 
$\hat{\bf S}_{i}^2|m_i\rangle = s(s+1) |m_i\rangle$ and
$\hat{S^z_i}|m_i\rangle=m_i|m_i\rangle$ where $|m_i| \leq s$.
Within this basis it is easy to distinguish between the subspaces
with different values of the projection of the total spin on the $z$-axis, $M$. 
In fact, in order to restrict to a particular $M$ sector, one has to use
only the states of the basis, $|x\rangle$, such that $\sum_{i}m_{i}=M$.
In addition, sorting the basis in order to group together the states with the same $M$,
the Hamiltonian matrix assumes a simpler block-diagonal form, i.e., a block
for each $M$ sector. Let us  restrict therefore to a particular $M$ subspace.

The first step of the proof is to perform a unitary transformation,
\begin{equation}
\calhat{U}^\dagger  = 
\exp{\Big[-i \pi \sum_{i \in {\rm B}} (s+\hat{S}_{i}^z) \Big]}~,
\label{eq.rotpi}
\end{equation}
whose physical meaning is to flip the quantization axis on the B 
sublattice. This defines a spatially varying reference frame 
pointing along the local N\'eel direction [Fig.~\ref{fig.neelj1j2}~(a)].  
The transformed Hamiltonian then results 
\begin{equation}
\calhat{U}^\dagger\calhat{H}\,\calhat{U}=\calhat{H}_{d}+\calhat{H}_{o}, 
\end{equation}
where the diagonal part,
\begin{equation}
\calhat{H}_{d}=  \sum_{(i,j)} J_{ij}
\hat{S}^z_{i} \hat{S}^z_{j}~,
\end{equation}
is invariant in the new representation, while the off-diagonal spin-flip term,
\begin{equation}
\calhat{H}_o = - \frac{1}{2} \sum_{(i,j)} J_{ij}
\big(\hat{S}^+_{i} \hat{S}^-_{j}+ h.c. \big)~,
\end{equation}
acquires instead an overall minus sign.
Therefore, in this representation, the Hamiltonian has non-positive off-diagonal 
matrix elements. In this case, using the hypothesis that the 
Hamiltonian cannot be split into sets of noninteracting spins 
one can demonstrate (Perron-Frobenius theorem\cite{perron})
that the ground-state expansion over the chosen basis 
$|\tilde{\psi}_{M}\rangle=\sum_{x}f_x|x\rangle$
has non-vanishing positive  amplitudes, i.e., $f_x>0$.

The latter result has two important consequences:
in each $M$ subspace the ground state of the Hamiltonian (\ref{eq.heis})
{\em i)} is {\em non-degenerate}, 
and {\em ii)} obeys, in the original representation, the well-known 
{\em Marshall-Peierls sign rule}, i.e.,
\begin{equation}
|\psi_M\rangle=\calhat{U}\,|\tilde{\psi}_{M}\rangle=
\sum_{x} e^{i \pi N(x)}f_{x}|x\rangle=
\sum_{x} (-1)^{N(x)}f_{x}|x\rangle~, 
\label{eq.mpsr}
\end{equation} 
where $N(x)=\sum_{i\in {\rm B}}(s+m_i)$, and the sum
is restricted to the configurations $|x\rangle$ such that $\sum_{i} m_{i}=M$.
Notice that the unessential constant term $s$ in the definition of $\calhat{U}$ 
have allowed us to write a real ground-state wavefunction.
In  particular, for $s=1/2$, $N(x)$ is simply the number of spins
up on the B sublattice, say $N_\uparrow(x)$. In this case, the ground-state projections 
on two Ising configurations differing for a single spin flip have opposite signs.
Notice that the rotational invariance of the Hamiltonian
has never been used in the proof so that this result is valid in general even in presence
of an easy-plane anisotropy, i.e., for the more general XXZ Hamiltonian,
or in presence of an arbitrary magnetic field in the $z$ direction. 

In contrast, for frustrated spin systems, like the $J_1-J_2$ model
and the Heisenberg triangular antiferromagnet,
the same proof does not hold.
In fact, in both systems the off-diagonal part of the
Hamiltonian cannot be made non-positive defined 
by any known unitary transformation.
In the $J_1-J_2$ model, the unitary transformation (\ref{eq.rotpi})
does not change the signs of the next-nearest-neighbors spin-flip term.
In the triangular antiferromagnet, under the transformation corresponding
to the one in Eq.~(\ref{eq.rotpi}) 
\begin{equation}
\calhat{U}^\dagger  = \exp{\Big[- \frac{2\pi i}{3}
\Big( \sum_{i \in {\rm B}} \hat{S}_{i}^z
- \sum_{i \in {\rm C}} \hat{S}_{i}^z \Big) \Big]}
\label{eq.rot2pi3}
\end{equation}
where B and C label two of the three sublattices as shown in Fig.~\ref{fig.neeltaf}
\footnote{As the unitary transformation in Eq.~(\ref{eq.rotpi}) 
this mapping defines a spatially varying coordinate system pointing
along the local N\'eel direction.}, 
the Hamiltonian reads 
\begin{eqnarray}
\calhat{U}^\dagger\calhat{H}\,\calhat{U}&=&-\frac{J}{4}
\sum_{\langle i,j \rangle} (\hat{S}^{+}_i \hat{S}^{-}_j+ h.c. )
+J\sum_{\langle i,j \rangle}\hat{S}^{ z}_i \hat{S}^{z}_j 
\nonumber\\
&+& 
i \sum_{\langle i,j \rangle} C_{i,j}(\hat{S}^{+}_i \hat{S}^{-}_j-h.c.)
\end{eqnarray}
with $C_{i,j}=\pm\sqrt{3}/4$.
The transformed Hamiltonian then displays an extra current-like term 
which is off-diagonal and has no definite sign.
Therefore for frustrated spin systems 
the Marshall-Peierls sign rule cannot be demonstrated and 
in general it does not hold exactly. 
In addition, the impossibility to find a unitary transformation
allowing us to map the Hamiltonian into an operator with non-positive defined
off-diagonal matrix elements has also dramatic consequences on the computability
of such frustrated systems with standard quantum Monte Carlo methods.
This is the origin of the so-called {\em sign-problem instability}.\cite{ss,sr}

However, as originally pointed out by Richter and co-workers,\cite{richter} 
the Marshall-Peierls 
sign-rule survives in the $J_1-J_2$ model in very good
approximation up to relatively large values of the frustration.
By means of the Lanczos exact diagonalization technique,
for $s=1/2$, $N=32$ and 36, we have calculated the weight of the states
satisfying the Marshall-Peierls sign-rule
in the expansion of the (normalized) exact ground state $|\psi_0\rangle$, namely:
\begin{equation}
\langle s \rangle = \sum_x |\psi_0(x)|^2 (-1)^{N_\uparrow(x)}{\rm sgn}\big[\psi_0(x)\big]~,
\end{equation}
with $\psi(x)=\langle x|\psi\rangle$ and the notations introduced in this
section.
Our results, shown in Tab.~\ref{tab.marshall}, put further evidence to the previous
findings of Ref.~\cite{richter} and indicate that the Marshall-Peierls
sign rule is verified almost exactly up to $J_2/J_1\simeq 0.3 \div 0.4$, even for
$N=36$. Moreover, even if the average sign $\langle s \rangle$ eventually
vanishes in the thermodynamic limit, its small size dependence
suggests that this property is likely to be conserved also for the lattice sizes
($N\simeq100$) presently accessible with the stochastic numerical techniques
used in this work.
A reasonable guess on the phases
of the exact ground state is in general very useful
to improve the efficiency of the approximated quantum
Monte Carlo techniques that have to be used in presence of the sign problem.\cite{ss,sr}

\begin{table}
\tcaption{Weight of the states satisfying the Marshall-Peierls sign-rule, $\langle s\rangle$,
in the ground state of the $J_1{-}J_2$ Heisenberg model.
Data are reported for $s=1/2$, $N=32$ and $N=36$, and various values of the frustration.}
\centerline{\footnotesize\smalllineskip
\begin{tabular}{c c c c c c}\\
\hline
$J_2/J_1$ &   0.1   & 0.2   &  0.3             & 0.4           &  0.5   \\
\hline
$N=32$    &   1     & 1     &  $\sim(1-10^{-8})$ & $\sim 0.9998$ &  0.973  \\
$N=36$    &   1     & 1     &  $\sim(1-10^{-8})$ & $\sim 0.9995$ &  0.961  \\
\hline\\
\end{tabular}}
\label{tab.marshall}
\end{table}

In the triangular case, instead, the phases
obtained by applying the operator (\ref{eq.rot2pi3})
to a state with positive-definite amplitudes on the Ising basis, as in 
Eq.~(\ref{eq.mpsr}), are very far from being exact, 
especially in the spin-isotropic limit (see Sec.~3.2.1).

\subsubsection{Ordering of the energy levels}

It follows from the spin rotational invariance of the Hamiltonian (\ref{eq.heis})
that each energy level, $E(S)$,  belonging to the total spin $S$ subspace
is $(2S+1)$-fold degenerate: in any $S$ sector there is a degenerate level 
for each value of $M$ in the range $-S\leq M \leq S$. 
Therefore, in a given $M$ subspace every energy eigenstate with  
total spin $S\ge|M|$ must be contained.  
In the hypothesis stated above, it is possible to prove
that the lowest energy in each $M$ subspace belongs to $S=M$, i.e.,
it has the minimum total spin allowed.
 
In order to prove this result we will show
that the ground state of $\calhat{H}$ in an $M$ subspace
is not orthogonal to the ground state of a rotational invariant soluble
Hamiltonian, which is known to belong to the $S=M$ sector. Therefore,
so does the former since two eigenfunctions having different quantum numbers 
are in general orthogonal.
Let us consider the infinite-range Heisenberg Hamiltonian on a bipartite lattice
\begin{equation}
\calhat{H}_\infty=J\sum_{i\in A,j\in B} \hat{{\bf {S}}}_{i} \cdot \hat{{\bf {S}}}_{j}~,
\label{eq.infran}
\end{equation}
with $J$ positive constant. This Hamiltonian is rotationally invariant
and exactly soluble since it is equivalent to a two spin problem:
\begin{equation}
\calhat{H}_\infty=J\big(\hat{{\bf S}}^2 - \hat{{\bf S}}_A^2 - \hat{{\bf S}}_B^2\big)
\end{equation}
where $\hat{{\bf S}}_A^2$ and $\hat{{\bf S}}_B^2$ are the total spin squared on the
$A$ and $B$ sublattices, respectively.
The eigenvalues of this special Hamiltonian are
\begin{equation}
E_{\infty}(S)= \frac{J}{2}\big[S(S+1)-S_A(S_A+1)-S_B(S_B+1)\big]~,
\end{equation}
and are monotonically increasing with the total spin $S$. Then,
the ground state of $\calhat{H}_\infty$, in each $M$ subspace, has $S=M$
total spin.
Moreover, both $\calhat{H}$ and $\calhat{H}_\infty$ satisfy the 
requirements for the Marshall-Peierls sign
rule. Hence, their ground state in any $M$ sector are not orthogonal
since their overlap involves the sum of positive numbers. 
It follows that they have necessarily the same total spin quantum number $S=M$.
Therefore in a given $M$ subspace the lowest energy of the Heisenberg Hamiltonian 
has the minimum total spin allowed.

This implies in turn that $E(S)<E(S+1)$.
In fact, among the degenerate eigenfunctions
with $E(S+1)$, there is a representative in the $M=S$ subspace.
The latter has not the minimum total spin allowed for that subspace
and therefore it has an energy higher than $E(S)$.
This proves that the energy levels of the Heisenberg antiferromagnet
(\ref{eq.heis}) {\em increase monotonically with the total spin} and, in particular, that
{\em the absolute ground state is a singlet} and {\em non-degenerate} (Lieb-Mattis property).
\footnote{In fact, having $S=0$, it has only a representative in the $M=0$ subspace}

The above proof on the ordering of the energy levels 
is a direct consequence of the Marshall-Peierls sign rule 
and therefore it breaks in presence of frustration.
However the Lieb-Mattis property 
turns out to be verified even for frustrated spin systems. In particular, 
for symmetry reasons, the ground state on any {\em finite size} 
of the spin-isotropic Heisenberg antiferromagnet is believed to possess 
all the symmetries of the Hamiltonian and in particular to be a singlet, 
rotationally invariant, and non-degenerate.\cite{azaria}
Even if there is no rigorous theorem proving this property in
general, the latter turns out to be true on a finite size whenever
the cluster is large enough, it has an even number of sites, and the
boundary conditions do no frustrate the antiferromagnetic long-range order.\cite{bernu}
In any case, however, these symmetry properties concern in general
the ground states {\em on finite sizes only}.
In the thermodynamic limit, the situation can change drastically
if there is no gap in the excitation spectrum. In this case,
in fact, a family of excited states collapses onto the ground state
and may break its symmetric character. 
This will be illustrated in the following sections.

\subsection{Order parameters and susceptibilities}
\label{sec.opags}

A zero-temperature spontaneously broken symmetry occurs 
when the ground state has a lower degree of symmetry 
than the corresponding Hamiltonian.
In this case, one can define an {\em extensive} operator, $\hat{O}$, breaking
some symmetry of the Hamiltonian and such that the so-called
{\em order parameter}, {\em i.e.}, the ground-state expectation value  
$m=\langle\psi_0|\hat{O}|\psi_0\rangle/N$, has a finite value.
In general, whenever the symmetry-breaking operator $\hat{O}$ 
does not commute with the Hamiltonian,
the symmetry breaking can happen only in the thermodynamic limit.
In fact, in that case, the ground-state expectation value
of $\hat{O}$ is zero on {\em any finite size} by symmetry.
This will be the case for the symmetry-breaking operators
considered in this paper.

The occurrence of a spontaneously broken symmetry can be detected
by adding to the Hamiltonian $\calhat{H}$ an {\em ordering field} $\delta$:
\begin{equation}
\calhat{H}_\delta=\calhat{H} - \delta\hat{O}.
\end{equation}
Since on a finite size the ground-state expectation value of $\hat{O}$
vanishes for $\delta=0$, the ground-state energy per site
has corrections proportional to $\delta^2$,
\begin{equation}
e(\delta)\simeq e_0-\frac{1}{2}\chi_O \delta^2 ~,
\end{equation}
$\chi_O$ being the (positive-definite) generalized susceptibility associated to the operator
$\hat{O}$, namely:
\begin{equation}
\chi_O = \frac{2}{N} \langle \psi_{0}| \hat{O} (E_{0}-\calhat{H})^{-1} \hat{O} | \psi_{0} \rangle~,
\label{eq.chi}
\end{equation}
where $E_{0}$ is the ground-state energy of $\calhat{H}$.

If symmetry breaking occurs in the thermodynamic limit then
\begin{equation}
\lim_{\delta\to0}\lim_{N\to\infty} \frac{1}{N}\langle\psi_0|\hat{O}|\psi_0\rangle=m\ne0,
\label{eq.opar}
\end{equation}
and the finite-size susceptibility has to diverge with the system size.
In fact, by the Hellmann-Feynman theorem, the ground-state expectation value
of $\hat{O}$ at finite field is $\langle \hat{O} \rangle_\delta/N = -de(\delta)/d\delta$,
so that, if symmetry breaking occurs in the thermodynamic limit,
an infinitesimal field $\delta$ must give a finite
$\langle \hat{O} \rangle_{\delta}/N \sim \chi_O \delta$
implying that the susceptibility has to diverge
in the thermodynamic limit. Moreover, it is possible
to show that the finite-size susceptibility must diverge
at least as the volume squared $N^2$. This will be proven in the next section.

\subsubsection{Exact bounds on the susceptibilities}
\label{sec.ebots}

Since on a finite size $\langle\psi_0|\hat{O}|\psi_0\rangle = 0$, it is 
convenient to introduce as the order parameter 
the quantity $p=\sqrt{\langle\psi_0|\hat{O}^2|\psi_0\rangle/N^2}$.
The latter is finite in general on any finite size and extrapolate to a finite value
in the thermodynamic limit in presence of long-range order, i.e., whenever
$m$, given by Eq.~(\ref{eq.opar}), is finite.
In this section we will show that, whenever symmetry breaking occurs in the thermodynamic limit,
the corresponding susceptibility must diverge as $N\to \infty$ and, in particular, it is bounded
from below by the order parameter times the system volume squared, 
namely $\chi_O >  const\,p^4 N^2$.

Let us define the following decomposition:
\begin{equation}
p^2= \frac{1}{N^2} \langle \psi_0| \hat{O}^2 | \psi_0\rangle =
 \frac{1}{N^2} \sum_{n\ne 0} |\langle \psi_0|\hat{O}|\psi_n\rangle|^2 =  \frac{1}{N} 
\int\hspace{-1mm}d\omega\,S(\omega)~
\label{eq.spectral}
\end{equation}
with
\begin{equation}
S(\omega) = \frac{1}{N} \sum_{n\ne0} |\langle \psi_0|\hat{O}|\psi_n\rangle|^2 \delta(\omega-\omega_{n})~,
\end{equation}
where we have introduced a complete set of eigenstates of the 
Hamiltonian $|\psi_n \rangle$ with eigenvalues $E_n$, we have used the symmetry
of the ground state (i.e., $\langle \psi_0 |\hat{O}|\psi_0\rangle=0$) and set 
$\omega_{n}=E_n-E_0$. 
By the Cauchy-Schwartz inequality we have:
\begin{eqnarray}
\int\hspace{-1mm}d\omega\,S(\omega)&=&\int d\omega\, \omega^{1/2} S(\omega)^{1/2} 
\omega^{-1/2} S(\omega)^{1/2} \nonumber \\
&\leq& \Big[ \int\hspace{-1mm}d\omega\,\omega S(\omega) \int\hspace{-1mm}d\omega\,\omega^{-1} S(\omega)\Big]^{1/2}~.
\label{eq.schwartz}
\end{eqnarray}
Now, 
\begin{equation}
\int\hspace{-1mm}d\omega\,\omega^{-1} S(\omega)=\frac{1}{N} \sum_{n\ne 0} \frac{1}{\omega_{n}}
|\langle \psi_0|\hat{O}|\psi_n\rangle|^2 \equiv \frac{\chi_O}{2}~,
\label{eq.defchi}
\end{equation}
where $\chi_O$ by Eq.~(\ref{eq.chi}) is the susceptibility associated to the
operator $\hat{O}$.
In addition it is straightforward to show that 
\begin{equation}
\int\hspace{-1mm} d\omega\,\omega S(\omega) = \frac{1}{N} \sum_{n\ne 0} \omega_{n}
|\langle \psi_0|\hat{O}|\psi_n\rangle|^2
=\frac{1}{2N} \langle[\hat{O},[\calhat{H},\hat{O}]]\rangle\equiv \frac{f_0}{2},
\end{equation}
so that, using Eqs.~(\ref{eq.schwartz}) and (\ref{eq.defchi}), we have
\begin{equation}
p^2 = \frac{1}{N}\int\hspace{-1mm} d\omega\,S(\omega) \leq \frac{1}{2N} \sqrt{\chi_O f_0}~.
\end{equation}
Hence, we have obtained the following lower bound for the susceptibility:
\begin{equation}
\chi_O\ge \frac{4 p^4}{f_0} N^2.
\label{eq.chidiv}
\end{equation}
Therefore, if the order parameter is finite in the thermodynamic limit,
the susceptibility must diverge at least as the volume squared, provided
$f_0$ is a constant. This happens whenever the commutator of
$\calhat{H}$ and $\hat{O}$ is an extensive quantity as
it is the case for the magnetization in the Heisenberg model and  
for all the symmetry-breaking operators treated in this work.

Moreover it is also possible to construct an upper bound for the
generalized susceptibility $\chi_O$ associated to a symmetry-breaking operator $\hat{O}$.
In fact, using Eqs.~(\ref{eq.defchi}) and (\ref{eq.spectral}), we have
\begin{equation}
\frac{\chi_O}{2}=\frac{1}{N}\sum_{n\ne 0}\frac{1}{\omega_{n}}|\langle \psi_0|\hat{O}|\psi_n\rangle|^2
\leq \frac{1}{\Delta} \int\hspace{-1mm} d\omega\,S(\omega) = \frac{N p^2}{\Delta}~,
\label{eq.boundchi}
\end{equation}
where $\Delta$ is the energy gap between the ground state and the first excitation.
An energy gap in the thermodynamic excitation spectrum is therefore 
incompatible with a spontaneously broken symmetry.
The physical meaning of this quite general result is the following: in presence
of a gap in the excitation spectrum, the ground state, which has generally all the symmetries
of the Hamiltonian on {\em any finite size}, has clearly no mean to develop a spontaneously
broken symmetry in the thermodynamic limit. In this case, the susceptibility is bounded
[Eq.~(\ref{eq.boundchi})]. 
In contrast, in presence of a gapless excitation spectrum, 
a family (or a {\em  tower}) of excited states can collapse 
in the thermodynamic limit onto the ground state
and can break its symmetric character. In fact, these states acquire
in general a phase factor under some operation of the symmetry group 
of the Hamiltonian and they can give rise to a symmetry-broken superposition.
Whenever this happens, the related susceptibility
must diverge as the volume squared [Eq.~(\ref{eq.chidiv})]. In this case from Eqs.~(\ref{eq.chidiv})
and (\ref{eq.boundchi}) we get a remarkable relation for the size-dependence of the spin gap, namely
\begin{equation}
\Delta \leq \frac{f_0}{2p^2} N^{-1}~.
\label{eq.sizegap}
\end{equation}

The mechanism underlying the spontaneous symmetry breaking 
leading to the onset of long-range N\'eel order in the thermodynamic ground state 
of a quantum antiferromagnet will be discussed in more detail in the following section.

\subsection{N\'eel order and Anderson's towers of states}
\label{sec.atos}

As we have already noticed, the occurrence of a spontaneous symmetry breaking
in the thermodynamic ground state, can be evidenced from the structure of the finite-size
energy spectrum.

On any finite size, the ground state of a quantum antiferromagnet
is generally believed to be a singlet, rotationally invariant and non-degenerate,
i.e., non-magnetic.
This is rigorously stated by the Lieb-Mattis theorem only
for the Heisenberg square antiferromagnet but nonetheless it can be numerically 
verified on small clusters also in presence of frustration for
the triangular Heisenberg antiferromagnet\cite{bernu} and for the $J_1{-}J_{2}$
model itself.\cite{schultz} 
Therefore, as originally pointed out by Anderson,\cite{anderson3}
the spontaneously symmetry breaking mechanism necessarily involves
the low-energy portion of the excitation spectrum:  in particular,
a whole {\em tower of states} has to collapse in the thermodynamic limit
onto the ground state faster than the low-lying excitations 
involving a spatial modulation of the classical N\'eel state (the so-called
{\em magnons}).
In fact, since in general these states acquire phase factors under 
rotations in the spin space, they can
sum up to a nontrivial state in which the spins
point in a definite direction, giving rise to N\'eel-like long-range
order. 

In particular, it is well known\cite{anderson4}$^-$\cite{hasenfrantz}
that in this case the low-lying excited states of
energy $E(S)$ and spin $S$ are predicted to behave as
the spectrum of a free quantum rotator (or {\em quantum top}) as long as $S \ll \sqrt{N}$,
\begin{equation}
E(S) -E_0 = \frac{S(S+1)}{2IN}~,
\label{eq.rotator}
\end{equation}
where $E_0=E(0)$ is the energy of the ground state, $|\psi_0\rangle$, and
$I$ is known as the {\em momentum of inertia} per site and is an intensive quantity.

This equation, which is in agreement with the bound (\ref{eq.sizegap}),
can be justified in a semiclassical picture of the long-range ordered
ground state of a quantum antiferromagnet.\cite{bernu}
To this purpose, let us consider the nearest-neighbor Heisenberg antiferromagnet on the
square lattice, and separate in the Fourier transformed Hamiltonian the ${\bf k}=0$,
and ${\bf Q}=(\pi,\pi)$ contributions (i.e., the only ones allowed by the sublattice 
translation invariance of the classical N\'eel state) from the others:
\begin{equation}
\calhat{H}=\calhat{H}_0+\calhat{V}~,
\end{equation}
where
\begin{equation}
\calhat{H}_0 = \frac{4J}{N} ( \hat{\bf S}^2 - \hat{\bf S}^2_A - \hat{\bf S}^2_B)~,
\end{equation}
$\hat{\bf S}^2$ is the total spin square,
and $\hat{\bf S}^2_A$ and $\hat{\bf S}^2_B$ are the total spin square of the A and B 
sublattices, respectively;
\begin{equation}
\calhat{V}= 2 J \sum_{{\bf k}\ne 0,{\bf Q}} 
\gamma_{\bf k}\, \hat{\bf S}_{\bf k} \cdot \hat{\bf S}_{-{\bf k}}~,
\end{equation}
where $\hat{\bf S}_{\bf k}=1/\sqrt{N}\sum_i \hat{\bf S}_{i} \exp ({\bf k}\cdot {\bf r}_i)$,
${\bf r}_i$ is the position of the site $i$, and $\gamma_{\bf k}= (\cos k_x + \cos k_y)/2$.  
For each value of the total spin $S$ the lowest eigenstate of $\calhat{H}_0$ is the classical-like 
state fully polarized on each magnetic sublattice with energy:
\begin{equation}
E_0(S)=- \frac{1}{4}(N+4)+ \frac{4JS(S+1)}{N}~. 
\end{equation} 
This is in agreement with the quantum top law (\ref{eq.rotator}).
Of course, since $\hat{\bf S}^2_A$ and $\hat{\bf S}^2_B$ 
do not commute with $\calhat{V}$, such eigenstates are not eigenstates of $\calhat{H}$.
Nonetheless we can look at them as the first approximation to the low-lying 
excited states in each $S$ sector.  The perturbation $\calhat{V}$ dresses 
these classical-like states with quantum fluctuations decreasing the average 
value of the sublattice magnetization and lowering their energy towards 
the exact result. However, as long as the  energy scale of these states
is well separated by the low-lying excitations with ${\bf k}\ne 0, {\bf Q}$ 
such renormalizations are not expected to modify the behavior (\ref{eq.rotator}).
These excitations are known as {\em magnons} or {\em spin-waves} (see Sec.~3.1), 
and involve a spatial modulation of the classical N\'eel state. 
In the Heisenberg antiferromagnet the dispersion relation of the softest
magnons is linear in the wavevector so that, 
in two dimensions, they have an energy scaling as $1/\sqrt{N}$. 
This implies that the constraint on the value of $S$ for the validity 
of Eq.~(\ref{eq.rotator}) is $S \ll \sqrt{N}$. 
 
The quantum top law is similar to the definition of the uniform spin 
susceptibility.\cite{phdthesis}  
The latter, in fact, can be calculated, by taking {\em first}
the infinite-volume limit of the energy per site $e(m)=E(S)/N$ at fixed
magnetization $m=S/N$ and then letting $m\to 0$ in the expansion 
\begin{equation}
e(m)=e_0+\frac{m^2}{2\chi}~,
\end{equation}
which is quite similar to Eq.~(\ref{eq.rotator}).
However an identification between $I$ and $\chi$ is possible if the excitation
spectrum smoothly connects the low-energy portion which
corresponds to total spin $S\sim{\cal O}(1)$, with the regime of
macroscopic spin excitations: $S\sim mN$ (with $m \ll 1$).\cite{lavalle} 
This is an highly nontrivial
statement which is actually verified by the underlying low-energy effective model
of the quantum antiferromagnet, known as nonlinear 
$\sigma$ model (NL$\sigma$M).\cite{azaria,fisher,hasenfrantz,lavalle,chakravarty}
Therefore, the quantity
\begin{equation}
\frac{1}{\chi_S}=2 N\frac{E(S)-E_0}{S(S+1)}
\end{equation}
must approach the inverse of the spin susceptibility
for infinite size and for any spin excitation with $S\ll N$.

\subsection{Resonating Valence Bond states}
\label{sec.rvb}

A simple and clear picture of a non-magnetic ground state can be given in terms
of the so-called {\em Resonating Valence Bond} (RVB) wavefunctions.\cite{anderson1}
Here, for simplicity, we will restrict ourselves to the case 
of the spin-half square antiferromagnet even if these states can be used also 
for a generic value of the spin $s$ (Ref.~\cite{auerbach}) and different lattice 
geometries.\cite{sindzingre}

The RVB wavefunctions are linear superpositions of  valence bond states
in which each spin forms a singlet bond
with another spin on the opposite sublattice.
These states form in general a (overcomplete) basis
of the $S=0$ subspace 
so that any singlet wavefunction can be represented in terms of them.
In particular, with such RVB wavefunctions it is possible to describe either
a long-range ordered or a non-magnetic state by varying the bond-length distribution.\cite{auerbach} 
In order to clarify this point, let us consider the following class of RVB wavefunctions
for a system of $N$ spins:
\begin{equation}
|\psi_{RVB}\rangle = 
\sum_{i_\alpha\in A,j_\beta\in B} h(r_1)\dots h(r_n) 
\,\,\, (i_1\,j_1) \dots (i_n,j_n)~,
\label{eq.rvb2}
\end{equation}
where $n=N/2$, $r_m$ is the distance between the spins forming the $m^{th}$ singlet bond
$(i_m\,j_m)$, and $h(r_m)$ is a bond weight factor, function of its length.
The latter wavefunction has no long-range order whenever the short-ranged
bonds are the dominant one in the superposition (\ref{eq.rvb2}). 
More precisely, it has been numerically shown  by 
Liang, Doucot and  Anderson\cite{liang} 
that the RVB state (\ref{eq.rvb2}) has no long-range antiferromagnetic order  
for bonds that decay as rapidly as $h(r)\sim r^{-p}$,
with $p\ge 5$. Instead, if the weight functions decay slowly enough with the length of the bond,
then the RVB wavefunction has a finite value of the thermodynamic order parameter squared. 
In particular, if the weight factors $h(r)$ are independent 
of the bond length, the RVB wavefunction is the projection of the N\'eel state 
onto the singlet subspace.

\begin{figure}
\centerline{\psfig{bbllx=120pt,bblly=310pt,bburx=490pt,bbury=490pt,%
figure=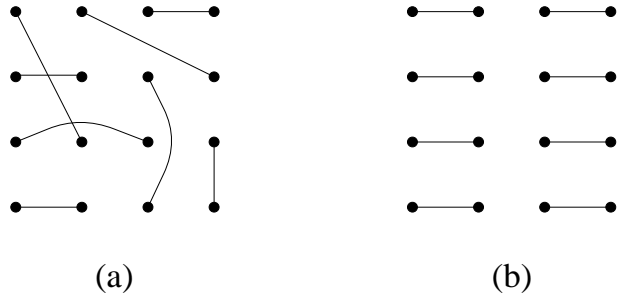,width=90mm,angle=0}}
\caption{\baselineskip .185in
An example of a spin liquid (a) and of a symmetry-broken (b) non-magnetic RVB state. 
Each stick represents a singlet bond.}\label{fig.rvb}
\end{figure}

Therefore, the simplest physical picture of a non-magnetic ground state
can be given in terms of a RVB wavefunction with short-ranged bonds. 
In addition, such bonds can be either homogeneously spatially distributed on the lattice, 
with short-range correlations among each other  ({\em spin liquid})
[Fig.~\ref{fig.rvb} (a)],
or they can break some symmetries of the Hamiltonian, with the dimers frozen
in some special patterns [Fig.~\ref{fig.rvb} (b)]. In Sec.~4.3
we will provide a possible example of the latter situation.

\section{The Triangular Heisenberg Antiferromagnet}
\label{chap.taf}

Historically the antiferromagnetic spin-1/2 Heisenberg model on the
triangular lattice was the first proposed Hamiltonian for a microscopic
realization of a non-magnetic ground state.\cite{anderson1,fazekas}
This is due to the fact that in this system the usual antiferromagnetic alignment 
between spins is hindered by the geometry of the lattice so that the minimum energy
configuration, the $2\pi/3$ N\'eel state (Fig.~\ref{fig.neeltaf}), 
has an energy twice larger than that on the square lattice and therefore is far less stable. 

However, despite a strong theoretical effort, 
the question of whether the combined effect of frustration and quantum
fluctuations in the ground state of the triangular Heisenberg antiferromagnet
favors a disordered RVB state 
or long-range N\'eel type order has been under 
debate for many years.\cite{bernu,azaria,sindzingre,oguchi1}$^-$\cite{manuel}

Spin-wave calculations\cite{jolicoeur,miyake} 
predict an important reduction (by about one-half)
of the sublattice magnetization by quantum fluctuations.
In addition, perturbation theory,\cite{oguchi2} series expansions,\cite{singh1}
and high-temperature calculations\cite{elstner}  suggest
that the spin-wave calculations possibly underestimate the
renormalization of the order parameter, but do not come to a definite conclusion about
the nature of ground state.
From the numerical point of view, exact diagonalization (ED) 
results,\cite{bernu,oguchi1,nishimori}$^-$\cite{leung}
which are limited to small lattice sizes, have
been interpreted both against\cite{leung} and in favor\cite{bernu} of
the presence of long-range N\'eel order in the thermodynamic ground state. 
In any case, this approach, combined with the 
careful analysis of the symmetry properties of the low-energy excited states
proposed by Bernu and co-workers,\cite{bernu} have provided  very important evidence
pointing towards a magnetic ground state:
the spectra of the lowest energy levels order with increasing total spin,
a reminiscence of the Lieb-Mattis theorem (see Sec.~2.1) for
bipartite lattices, and are consistent with the symmetry of the
classical order parameter. However, these very clear ED results 
cannot rule out that for large sizes quantum fluctuations could drive 
the system into a non-magnetic phase and therefore cannot be considered as conclusive.
In addition, standard stochastic numerical
methods, which usually allow one to handle large samples, clash with the sign problem
numerical instability so that a definite answer on the ground-state properties 
of the triangular Heisenberg antiferromagnet has been lacking for many years.

In this section we will tackle the problem of the existence of long-range
N\'eel order in the ground state of the triangular Heisenberg antiferromagnet
using the finite-size spin-wave theory,\cite{zhong} ED, and
Green function Monte Carlo (GFMC) methods. 
In the first part we apply the finite-size spin-wave theory to
the triangular Heisenberg antiferromagnet, we then show how to construct within this framework
the low-lying excited states, and finally derive a simple spin-wave variational wavefunction. 
The good agreement between the ED results and the finite-size spin-wave theory
will support the reliability of the
spin-wave expansion in describing not only the ground-state properties
but also the low-energy spin excitations of the Heisenberg model even in presence of frustration.
The second part will deal with the quantum Monte Carlo results. 
Data for the spin gap and for the antiferromagnetic order parameter will
be presented for fairly large system sizes (up to 144 sites), providing a robust evidence 
for a gapless excitation spectrum and for the existence of long-range N\'eel order
in the thermodynamic ground state of the model.

\subsection{Finite-size spin-wave theory}
\label{sec.fsswt}

Several attempts to generalize spin-wave theory to finite sizes can be found in the
literature.\cite{zhong,takahashi,hirsch} Here we will follow the
method proposed by Zhong and Sorella in Ref.~\cite{zhong} which allows one to deal with
finite clusters avoiding the spurious Goldstone modes divergences in a straightforward
way, and, in particular, without imposing any {\em ad hoc} holonomic
constraint on the sublattice magnetization.\cite{takahashi,hirsch}

\subsubsection{Application to the triangular antiferromagnet}

Assuming the classical ${\bf Q}=(4\pi/3,0)$ magnetic structure 
lying in the $xy$ plane, the first step of the derivation is to apply the unitary 
transformation given by Eq.~(\ref{eq.rot2pi3}), 
which defines a spatially varying coordinate system 
($x^\prime y^\prime z^\prime$) in such a way that the $x^\prime$-axis points on each site
along the local N\'eel direction. 
The transformed Heisenberg Hamiltonian reads:
\begin{eqnarray}
\calhat{U}^\dagger\calhat{H}\,\calhat{U}&=&J\sum_{\langle i,j \rangle} 
\Big[ \cos\left({\bf Q} \cdot {\bf r}_{i,j}\right) 
(\hat{S}^{x^\prime}_i \hat{S}^{x^\prime}_j+\hat{S}^{y^\prime}_i \hat{S}^{y^\prime}_j) 
\nonumber\\
&+& \sin\left({\bf Q}\cdot {\bf r}_{i,j}\right)
(\hat{S}^{x^\prime}_i \hat{S}^{y^\prime}_j-\hat{S}^{y^\prime}_i \hat{S}^{x^\prime}_j)
+\hat{S}^{ z^\prime}_i \hat{S}^{z^\prime}_j \Big]
\end{eqnarray}
where $J$ is the (positive) exchange constant between nearest 
neighbors, the indices $i,j$ label the points ${\bf r}_{i}$
and ${\bf r}_{j}$ on the $N$-site triangular lattice, ${\bf r}_{i,j}= {\bf r}_{i}-{\bf r}_{j}$,
and the quantum spin operators satisfy $|{\bf \hat{S}}_{i}|^2=s(s+1)$. 
In the new reference frame the spins in the classical configuration
are {\em ferromagnetically} aligned so that, 
using Holstein-Primakoff transformation for spin operators to order
$1/s$,
\begin{equation}
\hat{S}^{x^\prime}_{i} = s-\hat{a}^{\dagger}_i \hat{a}_i \,\,\,\,\,
\hat{S}^{y^\prime}_i = \sqrt{\frac{s}{2}}(\hat{a}^{\dagger}_i+\hat{a}_i)\,\,\,\,\,
\hat{S}^{z^\prime}_i = {\it i}\sqrt{\frac{s}{2}}(\hat{a}^{\dagger}_i-\hat{a}_i)~,
\end{equation}
being $\hat{a}$ and $\hat{a}^{\dagger}$ the canonical creation 
and destruction Bose operators, after some algebra the Fourier transformed Hamiltonian results: 
\begin{equation}
\label{eq.HSW}
\calhat{H}_{\rm SW}=E_{cl} +
3Js\sum_{\bf k}\Big[ A_{\bf k} \hat{a}^{\dagger}_{\bf k}\hat{a}_{\bf k}
+ \frac{1}{2}B_{\bf k}
(\hat{a}^{\dagger}_{\bf k}\hat{a}^{\dagger}_{-\bf k}+\hat{a}_{\bf k}\hat{a}_{-\bf k})\Big]
\end{equation}
where $E_{cl}=-3Js^2N/2$ is the classical ground-state energy, 
\begin{equation}
A_{\bf k}=1+\gamma_{\bf k}/2~,\:\: B_{\bf k}=-3~\gamma_{\bf k}/2~,
\label{eq.akbktr}
\end{equation} 
$\gamma_{\bf k} {=}2\left[ \cos\,(k_x)\,+2\,\cos\,(k_x/2)
\cos\,(\sqrt3\,k_y/2)\right]/z$, ${\bf k}$ is
a vector varying in the first Brillouin zone of the lattice, and $z=6$ is
the coordination number.
The Hamiltonian
$\calhat{H}_{\rm SW}$, can be diagonalized for ${\bf k} \neq 0, \pm {\bf Q}$ 
introducing the well-known Bogoliubov transformation, 
$\hat{a}_{\bf k}=u_{\bf k}\hat{\alpha}_{\bf k}+v_{\bf k}\hat{\alpha}^{\dagger}_{-{\bf k}}$,
with
\begin{equation}
u_{\bf k}=
\left(\frac{A_{\bf k}+
\epsilon_{\bf k}}{2\epsilon_{\bf k}}\right)^{1/2},\,\,\,
v_{\bf k}=-{\rm sgn}(B_{\bf k}) \left(\frac{A_{\bf k}-
\epsilon_{\bf k}}{2\epsilon_{\bf k}}\right)^{1/2}~,
\end{equation}
where $\epsilon_{\bf k}=\sqrt{A^2_{\bf k}-B^2_{\bf k}}$ is the  
spin-wave dispersion relation. This diagonalization leads to: 
\begin{equation}
\label{eq.diag}
{\cal H}_{\rm SW}^0=E_{\it cl}+  
\frac{3Js}{2}\hspace{-2mm}\sum_{{\bf k} \neq 0, \pm {\bf Q}}
(\epsilon_{\bf k}-A_{\bf k})+
\frac{3Js}{2}\hspace{-2mm}\sum_{{\bf k} \neq 0, \pm {\bf Q}}
\epsilon_{\bf k}
 (\hat{\alpha}^{\dagger}_{\bf k}\hat{\alpha}_{\bf k}~+
 \hat{\alpha}^{\dagger}_{\bf -k}\hat{\alpha}_{\bf -k}~).
\end{equation}

The Goldstone modes at ${\bf k}={\bf 0}$ and ${\bf k}= {\pm \bf Q}$ 
instead are singular, and cannot be diagonalized 
with a Bogoliubov transformation. 
For infinite systems such modes 
do not contribute to the integrals in Eq.~(\ref{eq.diag}), 
but in the finite-size case they are important and they must 
be treated separately. 
 By defining the following Hermitian operators 
\begin{eqnarray}
\hat{Q}_x&=&\frac{{\it i}}{2}(\hat{a}^{\dagger}_{\bf Q}
+\hat{a}_{-{\bf Q}}-\hat{a}_{\bf Q}-\hat{a}^{\dagger}_{-{\bf Q}})~,\nonumber \\
\hat{Q}_y&=&\frac{1}{2}(\hat{a}^{\dagger}_{\bf Q}
+\hat{a}_{-{\bf Q}}+\hat{a}_{\bf Q}+a^{\dagger}_{-{\bf Q}})~,\nonumber \\
\hat{Q}_z&=&i (\hat{a}^{\dagger}_0-\hat{a}_0) ~,
\end{eqnarray}
such that, $[\hat{Q}_{\alpha},\hat{Q}_{\beta}]=0$ and 
$[\hat{Q}_{\alpha},\calhat{H}_{\rm SW}]=0$ for $\alpha,\beta =x,y,z$,
the contribution of the singular modes, $\calhat{H}_{SM}$, in Eq.~(\ref{eq.HSW})
can be expressed in the form 
\begin{equation}
\calhat{H}_{SM}=-3JsA_{\bf 0}
+3Js\frac{A_{\bf 0}}{2}\left[\hat{Q}_x^{ 2}+\hat{Q}_y^{2}+\hat{Q}_z^{2}\right].
\end{equation}
Then, taking into account the fact that to the leading 
order in $1/s$, $\hat{Q}_{\alpha}=\hat{S}^{\alpha}\sqrt{2/Ns}$,
where $\hat{S}^{\alpha}$ are the components of the total spin, 
$\calhat{H}_{SM}$ may be also rewritten in the more physical form
\begin{equation}
\calhat{H}_{SM}=-3JsA_{\bf 0}
+3J\frac{A_{\bf 0}}{N}\left[(\hat{S}^x)^2+(\hat{S}^y)^2+(\hat{S}^z)^2\right],
\end{equation}
which clearly favors a singlet (${\bf S}^2=0$) ground state (for an even number of sites) 
being $A_{\bf 0}$ positive definite. 
This result is a reminiscence of 
the Lieb-Mattis property (see Sec.~2.1) 
which has not been demonstrated for non-bipartite lattices. 
Actually, a similar result is obtained by solving exactly
the three Fourier components ${\bf k}={\bf 0},\pm {\bf Q}$ of the 
Heisenberg model;\cite{bernu} however, our treatment allows us to
construct  a formal expression for the spin-wave ground state  
on finite triangular lattices which keeps the correct singlet behavior.
In fact, starting from the usual spin-wave ground state, composed by
the $2\pi/3$ classical N\'eel order plus the 
zero point quantum fluctuations (i.e,  zero Bogoliubov quasiparticles),
\begin{equation}
|0\rangle=\prod_{{\bf k}\neq {\bf 0},\pm{\bf Q}} u^{-1}_{{\bf k}}
{\rm exp}{\Big[\frac{1}{2}\frac{v_{{\bf k}}}{u_{{\bf k}}} 
\hat{a}^{\dagger}_{\bf k}\hat{a}^{\dagger}_{-{\bf k}}\Big]}|F\rangle
\end{equation}
with $|F\rangle=\prod_i|S^{x^\prime}_i=s\rangle $, the corresponding
singlet wavefunction is obtained by projecting $|0\rangle$ onto 
the subspace $S=0$:
\begin{equation}
|\psi_{\rm SW}\rangle=\int^{\infty}_{-\infty}\hspace{-1mm}d\alpha 
\int^{\infty}_{-\infty}\hspace{-1mm}d\beta\int^{\infty}_{-\infty}
\hspace{-1mm}d\gamma~
e^{{\it i}\alpha Q_x+{\it i}\beta Q_y+{\it i}\gamma Q_z}|0\rangle
\label{eq.swwf}
\end{equation}
and reads
$|\psi_{\rm SW}\rangle\sim e^{(-\hat{a}^{\dagger}_{\bf Q}\hat{a}^{\dagger}_{- \bf Q}+
\frac{1}{2}\hat{a}^{\dagger}_{\bf 0}\hat{a}^{\dagger}_{\bf 0})} |0\rangle~$.
In particular the singular modes have no contribution
to the ground-state energy which
reads
\begin{equation}
E_{\rm SW} =E_{\it cl}+ \frac{3Js}{2}\sum_{{\bf k}} (\epsilon_{\bf k}-1)~,
\label{eq.esw}
\end{equation}
while the computation of the order parameter requires their remotion:
\begin{equation}
\hat{m}^{\dagger}=\sqrt{\langle (\hat{S}^{x^\prime}_i)^2\rangle}=s-\frac{1}{N} \sum_{{\bf k}\neq 
{\bf 0},\pm {\bf Q}}v^2_{\bf k}~.
\label{eq.msw}
\end{equation}

For $s=1/2$, the previous spin-wave calculation predicts a very good
quantitative agreement with exact results on small
clusters ($N \le 36$) of both ground-state energy and sublattice
magnetization.\cite{triangsw} The agreement is even more remarkable as far
as the low-lying excited states are concerned, as it will be shown in the following section.

\subsubsection{Low-energy spin-wave spectrum}
\label{ssec.lesws}

In this  section, we show how to construct the low-lying energy
spectra $E(S)$  for finite systems where $S$ represents the total spin.
So far, we have performed a standard spin-wave expansion whose relevant
quantum number is $s$. Thus, computing $E(S)$ is not straightforward 
and require a little more involved calculation.   
Following Lavalle, Sorella and Parola,\cite{lavalle} 
a magnetic field in the $z$-direction is 
added to stabilize  the desired total spin excitation $S$:
\begin{equation}
\calhat{H}^h_{\rm SW}=\calhat{H}_{\rm SW}-hs\sum_{i}\hat{S}_{i}^{z}.
\end{equation}
Classically, for magnetic fields not large enough to induce a spin-flop transition,
the new solution is the $2\pi/3$ N\'eel order canted by 
an angle $\theta$ along the direction of the field $h$. In order to develop a spin-wave 
calculation, a new rotation around $y^\prime$-axis is performed on the spin
operators and it can be proven that ${\cal H}^h_{\rm SW}$
takes the same form of Eq.~(\ref{eq.HSW}) with renormalized coefficients $A_{\bf k}$
and $B_{\bf k}$:
\begin{equation}
A^h_{\bf k}=1+\gamma_{\bf k}\left[\frac{1}{2}-
\frac{3}{2}(\frac{2h}{z3J})^2\right],\:\:
B^h_{\bf k}=-\frac{3}{2}\gamma_{\bf k} \left[1-(\frac{2h}{z3J})^2\right]~,
\end{equation}
being $2h/z3J=\sin\theta$. For $h\ne 0$
the only singular mode is ${\bf k}= 0$ (associated to the rotation invariance
in the $xy$ plane) 
and its contribution is given by
\begin{equation}
{\cal H}_{SM}=-\frac{3Js}{2}\;A^h_{{\bf 0}} 
+ 3J \frac{A^h_{{\bf 0}}}{N}\: (S^z-Ns\; \sin\theta)^2, 
\end{equation}
which now  favors  a value of $S^{z}$ 
consistent  with the applied field, at the classical level. 

The Hellmann-Feynman theorem relates the total spin $S=N \langle S^z_{i} \rangle$ of the
excitation to the magnetic field $h$ as it follows:
\begin{equation}
\langle S^z_{i} \rangle = -\frac{1}{Ns}\frac{\partial}{\partial h}E(h)
=
s\frac{2h}{z3J} \left[1+\frac{1}{2Ns} \sum_{{\bf k}\neq0} \gamma_{\bf k}
\sqrt{\frac{A^h_{\bf k}+B^h_{\bf k}} {A^h_{\bf k}-B^h_{\bf k}}}\right]
\label{eq.hft}
\end{equation}
where 
\begin{equation}
\label{eq.EH}
E(h)=E_{cl}-\frac{1}{2}(sh)^2\frac{2N}{3zJ}-3Js\frac{N}{2}+
\frac{3Js}{2}\sum_{\bf k}\epsilon^h_{\bf k}.
\end{equation}
is the spin-wave energy in presence of the field 
and $\epsilon_{\bf k}^h=\sqrt{(A^h_{\bf k})^2-(B^h_{\bf k})^2}$.
In particular, the first term in $(sh)^2$ in Eq.~(\ref{eq.EH}) gives
the classical uniform spin susceptibility $\chi_{cl}=1/9J$, 
while taking the whole expression the known spin-wave result\cite{chubukov} 
$\chi_{\rm SW}/\chi_{cl}=1-0.449/2s$ is recovered. 
Finally, as suggested by Lavalle and co-workers,\cite{lavalle} given the value $S$,
the corresponding values of $h$ and of $E(h)$ can be found 
with Eqs.~(\ref{eq.hft}) and (\ref{eq.EH}),
and the energy of the spin excitation $E(S)$ can be calculated,
at fixed $s$, with a Legendre transformation $E(S)=E(h)+hsS$.\cite{phdthesis}

\begin{figure}
\centerline{\psfig{bbllx=65pt,bblly=180pt,bburx=520pt,bbury=600pt,%
figure=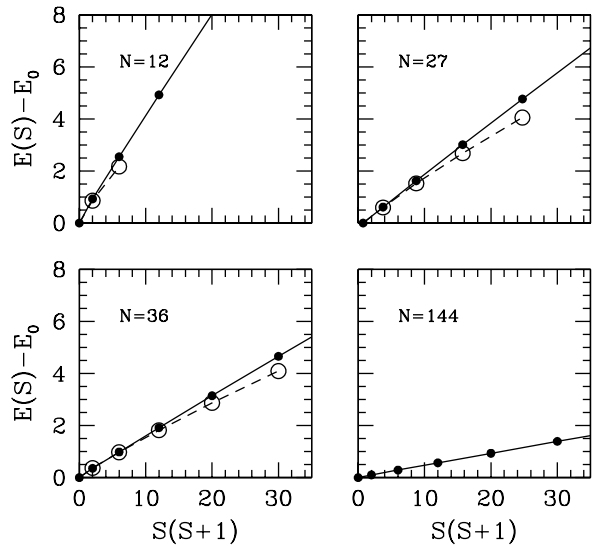,width=80mm,angle=0}}
\caption{Spin-wave (full dots and continuous line) and exact
(empty dots and dashed lines) low-energy spectra as a function
of $|{\bf S}^2|=S(S+1)$ for $N=12,27,36,144$ and $s=1/2$. }
\label{fig.taf1}
\end{figure}

As explained in Sec.~2.3, the occurrence of a symmetry breaking
in the ground state for $N\rightarrow \infty$
can be evidenced from the structure of the finite-size
energy spectra. In particular,
when long-range order is present in
the thermodynamic limit, the low-lying excited states of
energy $E(S)$ and spin $S$ are predicted to behave as
the spectrum of a free quantum rotator (\ref{eq.rotator})
as long as $S \ll \sqrt{N}$.
Actually, on the triangular lattice the quantum-top effective Hamiltonian
displays a correction due to the anisotropy of the susceptibility tensor.\cite{bernu} 
However, in the following we will consider only the leading contribution
(\ref{eq.rotator}) which depends on the perpendicular susceptibility. 
Fig.~\ref{fig.taf1} shows $E(S)$ vs $S(S+1)$ calculated within the spin-wave theory
compared with the exact diagonalization results of Bernu and 
co-workers.\cite{bernu} 
Remarkably the spin-wave theory turns out to be accurate
in reproducing the low-energy spectrum in the whole range of sizes.
Furthermore, we can extend our calculation to the thermodynamic
limit and observe easily the collapse of a macroscopic  number of
states with different $S$ to the ground state as
$N\rightarrow\infty$. This clearly gives rise to a  broken SU(2)
symmetry ground state, as expected within the spin-wave framework.

\begin{figure}
\centerline{\psfig{bbllx=65pt,bblly=160pt,bburx=505pt,bbury=690pt,%
figure=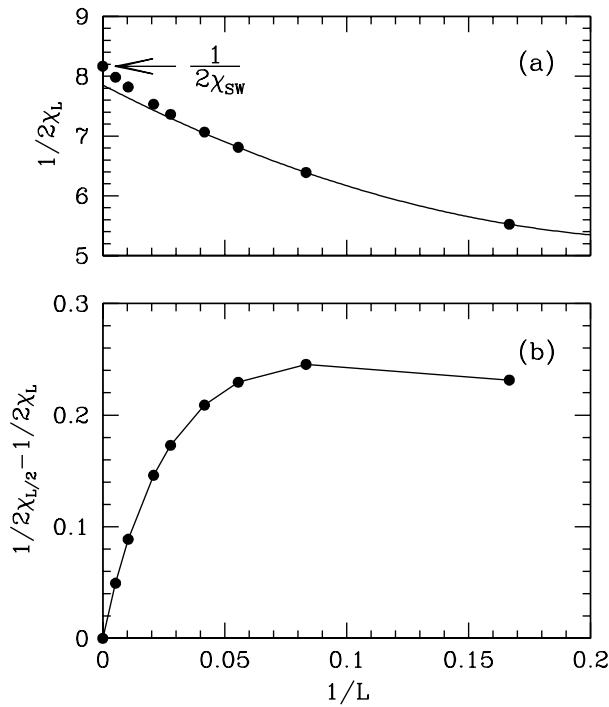,width=80mm,angle=0}}
\caption{Size dependence of 1/2$\chi_{L}$ (a)
and of $1/2\chi_{L/2}-1/2\chi_{L}$ (b) obtained according to
Eq.~(\ref{eq.inv}) using  the ($s=1/2$) spin-wave excitation spectra.
The continuous line is a quadratic fit for $L<18$ in (a)
and a guide for the eye in (b).}
\label{fig.taf2}
\end{figure}

In addition, as explained in Sec.~2.3,
whenever the quantum top law (\ref{eq.rotator}) is verified, the quantity
\begin{equation}
\left[ 2 \chi_{S} \right]^{-1} = N E(S)\left[S(S+1)\right]^{-1}~,
\label{eq.inv}
\end{equation}
should approach the physical inverse susceptibility $1/2\chi_{\rm SW}$
for infinite size and for any spin excitation $S\ll N$. This feature is
clearly present in the spin-wave theory and it is shown in Fig.~\ref{fig.taf2}~(a)
where the $1/2\chi_{S}$  is plotted for $S=L \equiv \sqrt{N}$ and approaches
the predicted value ($1/2\chi_{\rm SW}=8.167$), even if the correct asymptotic
scaling $1/2\chi_{L}\simeq 1/2\chi_{\rm SW}+ a/L + b/L^2$ turns out
to be satisfied only for very large sizes ($L\geq 36$). Such feature
is also shared by the Heisenberg antiferromagnet on the square lattice
where a similar spin-wave analysis has allowed the authors of Ref.~\cite{lavalle}
to account for the anomalous
finite-size spectrum resulting from an accurate quantum Monte Carlo calculation.
Furthermore, similarly to the latter case, a non-monotonic
behavior of $1/2\chi_{L/2}-1/2\chi_{L}$ [Fig.~\ref{fig.taf2}~(b)],
which should extrapolate to 0 as $1/L$ according to
the quantum top law,
persists also in presence of the frustration within the spin-wave approximation
and is likely to be a genuine feature of the Heisenberg model.

\subsubsection{Spin-wave variational wavefunctions}
\label{ssec.swvwf}

In a GFMC
calculation it is in general important to start from an accurate 
variational guess of the ground-state wavefunction. 
So far, many wavefunctions have been proposed in the
literature\cite{sindzingre,kalmeyer,huse,yang} and the lowest ground-state
energy estimation was obtained with the long-range ordered type.\cite{sindzingre,huse}

The simplest starting point for constructing a long-range ordered
wavefunction is of course the classical N\'eel state.
Since on finite-size the ground state is expected to be rotationally invariant,
the N\'eel state should be projected onto the ${\bf S}^{2}=0$
subspace. However it is in general numerically very difficult to perform the projection
onto a total spin subspace so that only the projection onto the subspace
with $S^{z}= 0$ (a quantum number of the states of the chosen basis)
is usually performed.
Quantum correction to this classical wavefunction
can be introduced by means of a Jastrow factor containing
all the two-spin correlations
\begin{equation}
|\psi_G\rangle= \hat{P}_0 \exp \Big(  {\eta \over 2} \sum_{i,j}
v(i-j) \hat{S}^{z}_{i} \hat{S}^{z}_{j} \Big) |N\rangle~,
\label{eq.lrowf}
\end{equation}
where $\hat{P}_0$ is the projector onto the $S^{z} = 0$ subspace.
Starting from the spin-wave ground state (\ref{eq.swwf}) it is possible to derive\cite{franjic}
a simple variational wavefunction which is both accurate and easily computable
also when used for importance sampling in a quantum Monte Carlo calculation.\cite{sr}
Such wavefunction is defined for any $s$ in the correct Hilbert space of the
spin operators and reduces for $s\to \infty$ to the spin wave form (\ref{eq.swwf}).

To this purpose let us consider the following variational wavefunction
\begin{equation}
|\tilde{\psi}_{G}\rangle = \hat{P}_0\, \exp{\Big[\frac{1}{s}\sum_{{\bf k}\ne 0} g_{\bf k}
\hat{S}^{z}_{\bf k} \hat{S}^{z}_{- \bf k}\Big]}|F\rangle~,
\label{eq.psit}
\end{equation}
where $\hat{S}^z_{\bf k} = N^{-1/2} \sum_{j} e^{-i {\bf k}\cdot{\bf r}_{j}} \hat{S}^z_{j}$.
The spin-wave $s\to\infty$ limit of the wavefunction (\ref{eq.psit}) can be easily
carried out by means of a Hubbard-Stratonovich transformation\cite{franjic}
and leads to (neglecting an unessential normalization)
\begin{equation}
|\tilde{\psi}_{G}\rangle=\hat{P}_{0} \exp{\Big[-\frac{1}{2}\sum_{{\bf k}\neq 0}
\frac{g_{\bf k}}{1-g_{\bf k}} \hat{a}^{\dagger}_{\bf k} \hat{a}^{\dagger}_{-{\bf k}}  \Big]}|F\rangle.
\end{equation}
By requiring that $|\psi_{G}\rangle$ reduces to the spin-wave wavefunction (\ref{eq.swwf})
for $s \to \infty$ one obtains for $g_{\bf k}$
\begin{equation}
g_{\bf k}= \frac{v_{\bf k}}{v_{\bf k}-u_{\bf k}}
=1-\sqrt{\frac{1+2\gamma_{\bf k}}{1-\gamma_{\bf k}}}
\end{equation}
which is singular only for ${\bf k}=0$.
This analysis, for the more general XXZ Hamiltonian 
with an exchange easy-plane anisotropy $\alpha$,\cite{xxzqtaf} 
gives
\begin{equation}
g_{\bf k}=1-\sqrt{ \frac{1+2\alpha\gamma_ {\bf k}}{1-\gamma_{\bf k}}}~.
\end{equation}

In the original (unrotated) reference frame,
the N\'eel state $|N\rangle$ can be written in terms of
$|F\rangle=\prod_i|S^{x^\prime}_i= s \rangle$ by applying the
inverse of the unitary transformation (\ref{eq.rot2pi3})
\begin{equation}
|N\rangle = \calhat{U} |F\rangle = \sum_x {\cal U}(x) |x\rangle =
\sum_{x} \exp{\Big[+\frac{2\pi i}{3} \Big( \sum_{i \in {\rm B}} S_{i}^z
- \sum_{i \in {\rm C}} S_{i}^z \Big) \Big]} |x\rangle~,
\label{eq.neel}
\end{equation}
where $|x\rangle$ is an Ising spin configuration specified
by assigning the value of $S^z_i$ (or equivalently of $S_i^{z \prime}$) for
each site, and ${\cal U}(x)=\langle x|\calhat{U}|x\rangle$.
Then, introducing a variational parameter $\eta$ scaling the latter potential,
for $s=1/2$ and in the original spin representation, $|\psi_G\rangle$ assumes the very simple form
of Eq.~(\ref{eq.lrowf}), i.e.,
\begin{equation}
|\psi_G\rangle= \calhat{U} |\tilde{\psi}_G\rangle
= \sum_{x} {\cal U}(x) \exp{\Big[ \frac{\eta}{2} \sum_{i,j}
v(i-j)S_{i}^zS_{j}^z\Big]}|x\rangle~,
\end{equation}
where
$v(r)=1/N\sum_{{\bf q}\neq 0} e^{-i {\bf q}\cdot{\bf r}} g_{\bf q}$
and the summation is restricted only to the Ising configurations
with $\sum_{i}S^z_i=0$ to enforce the projection onto the $S^z=0$ subspace.
In contrast to the linear spin-wave ground state (\ref{eq.swwf}), which
does not satisfy the constraint $\langle\hat{a}^{\dagger}_i \hat{a}_i\rangle \leq 2s$,
the present variational wavefunction is defined in the correct Hilbert space
of the spin operators.

\subsection{Quantum Monte Carlo calculation}

\subsubsection{From Marshal-Peierls to Huse-Elser sign rule}
\label{ssec.wvf}

According to the Marshall theorem (see Sec.~2.1),
for the Heisenberg antiferromagnet on the square lattice, as well as  on any other bipartite
lattice, 
the classical part of the wavefunction by itself determines exactly the phases of the
ground state in the chosen basis.  For the triangular case, instead, the exact phases 
are unknown and the classical part is not enough to fix them correctly.
In particular, starting from a long-range ordered wavefunction of the form (\ref{eq.lrowf}),
Huse and Elser\cite{huse} introduced important three-spin correlation factors in the 
wavefunction
\begin{equation}
\hat{T} = \exp{\Big({\it i}\,\beta\sum_{\langle i,j,k\rangle}
\gamma_{ijk} \hat{S}_{i}^z \hat{S}_{j}^z \hat{S}_{k}^z \Big)}~,
\label{eq.triplet}
\end{equation}
defined by the coefficients $\gamma_{ijk}=0,\pm 1$, appropriately
chosen so as to preserve the symmetries of the classical N\'eel state, and  by  an overall
factor $\beta$. In particular the sum in Eq.~(\ref{eq.triplet}) runs over all
distinct triplets of sites $i,j,k$ where both $i$ and $k$ are nearest neighbors of
$j$, and $i$ and $k$ are next-nearest neighbors to one another.
The sign factor $\gamma_{ijk}=\gamma_{kji}=\pm 1$ is invariant under rigid translations and rotations
in real space by an angle of $2\pi/3$ of the three-spin cluster $i,j,k$, but changes
sign under rotations by $\pi/3$ or $\pi$.
The resulting wavefunction reads therefore:
\begin{equation}
|\psi_G\rangle=  \sum_x \Omega(x)
\exp \Big(  {\gamma \over 2} \sum_{i,j}
v(i-j) S^{z}_{i} S^{z}_{j} \Big) |x\rangle~,
\label{eq.wfhuse}
\end{equation}
with a phase factor given by
\begin{equation}
\Omega(x)=T(x) \exp{\Big[+ \frac{2\pi i}{3}
\Big( \sum_{i \in {\rm B}} S_{i}^z
- \sum_{i \in {\rm C}} S_{i}^z \Big) \Big]}
\end{equation}
where $T(x)=\langle x|\hat{T}|x\rangle$.
Finally, since the Hamiltonian is real,
a better variational wavefunction on a finite size
is obtained by taking the real part of Eq.~(\ref{eq.wfhuse}).

\begin{figure}
\centerline{\psfig{bbllx=45pt,bblly=255pt,bburx=505pt,bbury=690pt,%
figure=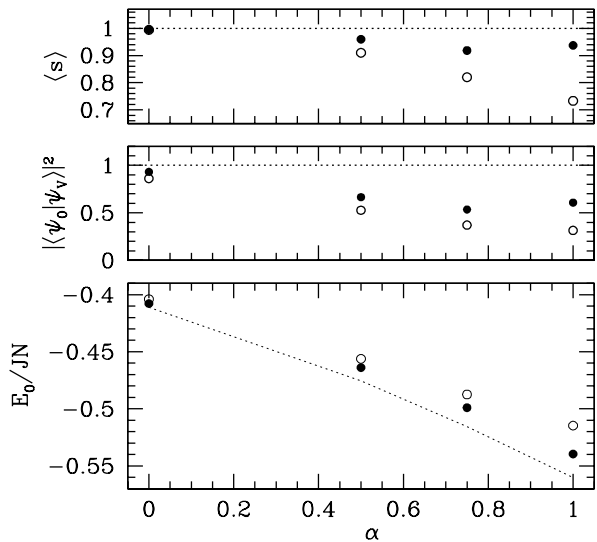,width=80mm,angle=0}}
\caption{Average sign, overlap square and ground-state energy  per site
obtained for $N=36$ using the variational wavefunction of Eq.~(\ref{eq.wfhuse}),
with (full dots) and without (empty dots) the triplet term
of Eq.~(\protect\ref{eq.triplet}),
as a function of the easy-plane anisotropy $\alpha$.
The  calculations were performed by summing exactly over all
the configurations and the dotted line connects the exact results.
}
\label{fig.taf3}
\end{figure}

Although the three-body
correlations of Eq.~(\ref{eq.triplet}) do not
provide the exact answer, they
allow us to adjust the signs of the wavefunction in a nontrivial
way without changing the underlying classical N\'eel order. 
In this respect it is useful to define an average sign of the variational
wavefunction relative to the normalized exact ground state  $|\psi_{0}\rangle$ as
\begin{equation}\langle s \rangle = \sum_{x} |\psi_{\rm 0}(x)|^2
{\rm sgn}\big[\psi_G(x)\psi_{\rm 0}(x)\big]~,
\end{equation}
with $\psi(x) = \langle x|\psi\rangle$.
We have compared the variational calculation
with the exact ground state  obtained by ED
on the $N = 36$ cluster. For completeness we have considered
the more general XXZ Hamiltonian with the exchange easy-plane anisotropy $\alpha$,
ranging from the XY case ($\alpha = 0$) to the standard spin-isotropic
case ($\alpha = 1$).
As shown in Fig.~\ref{fig.taf3},
the introduction of the three-body correlations
of Eq.~(\ref{eq.triplet}), although not providing the exact answer,
improves the overlap square of the variational wavefunction
on the true ground state and the accuracy of the variational estimate
of the ground-state energy as well.
In particular the average sign $\langle s \rangle$
is very much improved by the triplet term, particularly in the spin-isotropic limit
$\alpha \to 1$. This is crucial when the variational wavefunction
is used for importance sampling within the modifications of the
GFMC technique developed to handle the sign problem instability, 
like the FN and GFMCSR techniques.
In the following the wavefunction (\ref{eq.wfhuse}) 
will be used as the guiding wavefunction in
our quantum Monte Carlo calculations.

\subsubsection{The reconfiguration scheme}

\begin{figure}
\centerline{\psfig{bbllx=120pt,bblly=300pt,bburx=500pt,bbury=640pt,%
figure=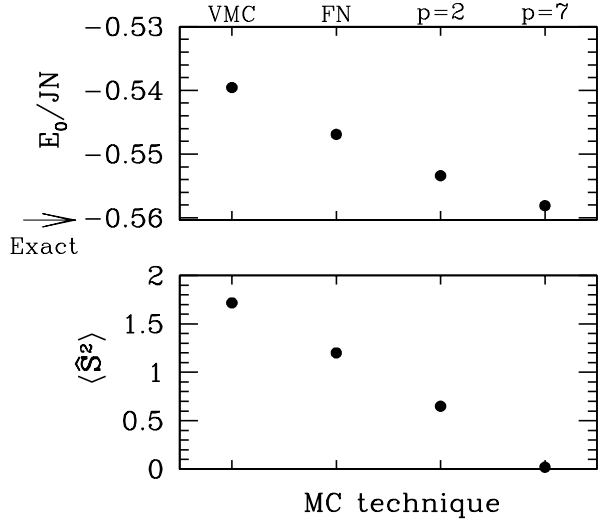,width=80mm,angle=0}}
\caption{Variational estimate (VMC) and mixed averages 
(FN, GFMCSR) of the ground-state energy  per site $E_{0}/JN$ and of the total spin
square $\langle \hat{\bf S}^2 \rangle$ for $N=36$.
GFMCSR data are obtained using the short-range correlation
functions generated by $\calhat{H}$ ($p=2$), and $\calhat{H}^2$
($p=7$) as explained in the text.
}
\label{fig.taf4}
\end{figure}

\begin{figure}
\centerline{\psfig{bbllx=60pt,bblly=330pt,bburx=540pt,bbury=560pt,%
figure=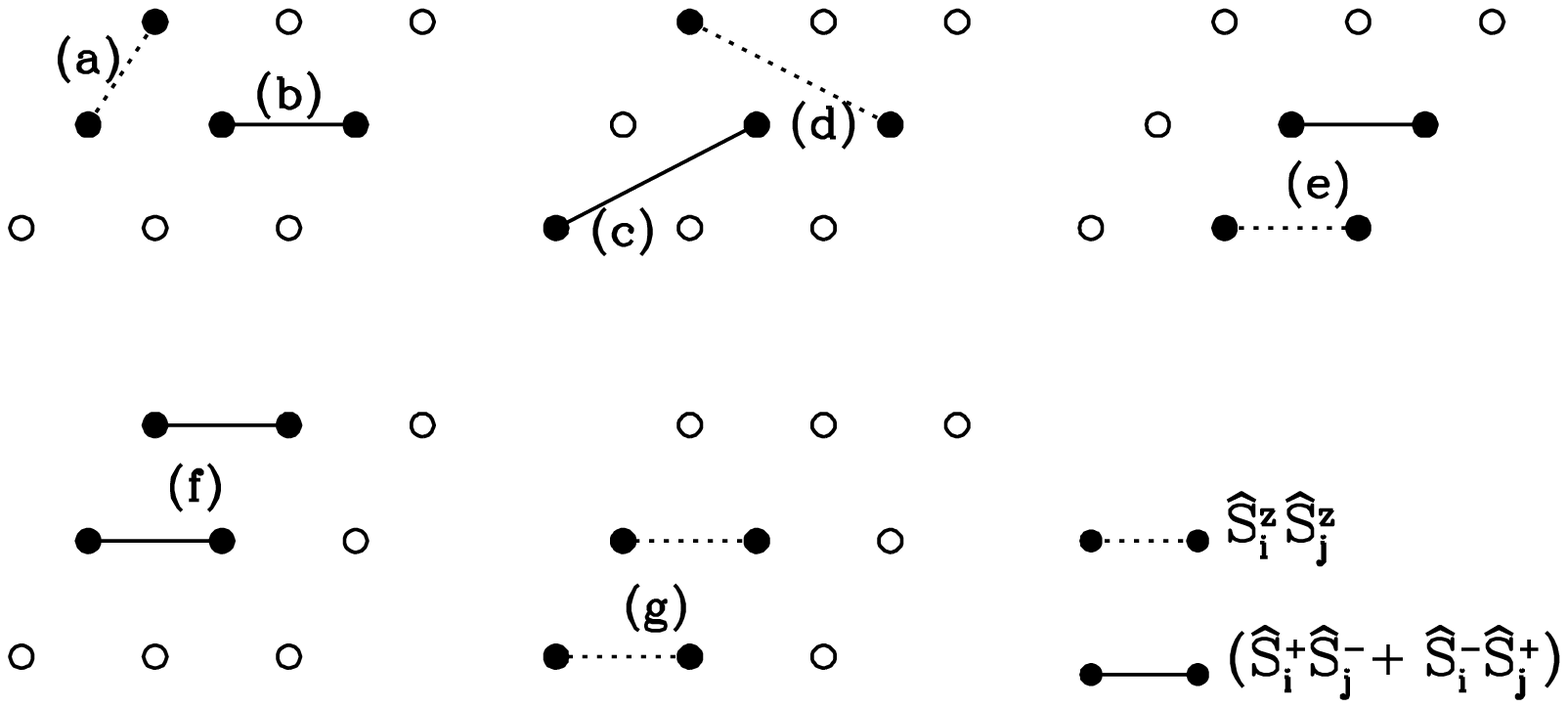,width=80mm,angle=0}}
\caption {\baselineskip .185in \label{fig.taf5}
Short range spin correlation functions generated by $\calhat{H}$ (a,b) and  $\calhat{H}^2$ (c-g).
}
\end{figure}

In order to study the ground-state properties we have used
the GFMCSR quantum Monte Carlo technique, which allows one to release
the FN approximation, in an approximate but controlled way. 
This systematic improvement introduced by the GFMCSR on the accuracy of the ground-state properties,
is illustrated in Fig.~\ref{fig.taf4}, where we display
a comparison between the estimates of the ground-state energy per site
and of total spin square, for the $N=36$,
obtained with the stochastic sampling of the variational wavefunction (\ref{eq.wfhuse}),
the FN and the GFMCSR techniques.
As explained in Refs.~\cite{ss,sr}, in the appropriate limit of large number of walkers and high
frequency of SR, the residual bias introduced by the GFMCSR depends  only
on the number $p$ of operators used to constrain the GFMC Markov process,
allowing  simulations without numerical instabilities.  In principle the exact answer
can be obtained, within statistical errors, provided $p$ equals
the huge Hilbert space dimension.
In practice it is necessary to work with small $p$ and an accurate selection
of physically relevant operators is therefore crucial. As can be easily expected,
the short-range correlation functions $\hat{S}^z_i\hat{S}^z_j$ and
$(\hat{S}^+_i\hat{S}^-_j{+}\hat{S}^-_i\hat{S}^+_j)$ contained
in the Hamiltonian ($p=2$) give a sizable improvement of the FN ground-state energy
when they are included in the SR procedure.
In order to be systematic, we have included in the SR also
the short-range correlations  generated by $\calhat{H}^2$ ($p=7$),
averaged over all spatial symmetries commuting
with the Hamiltonian.
These local  correlations (see Fig.~\ref{fig.taf5}) are particularly important to
obtain quite accurate and reliable estimates not only
of the ground-state energy but also of the mixed average\cite{ss,sr} of the total spin
square $\hat{\bf S}^2$.
In particular it is interesting that, starting
from a variational wavefunction with no definite spin, the spin rotational invariance of the
finite-size ground state is systematically recovered by means of the GFMCSR technique
(see lower panel of Fig.~\ref{fig.taf4}).

Having obtained an estimate for the ground-state energy, at least an order of magnitude
more accurate than our best variational guess, it appears possible
to obtain physical features, such as a gap in the spin spectrum,
that are not present  at the variational level.
For instance in the frustrated $J_1{-}J_2$ Heisenberg model (see Sec.~4.2), 
with the same technique and a similar accuracy,  a gap in the spin spectrum
is found in the thermodynamic limit, starting with a similar
ordered and therefore gapless variational wavefunction.

\subsubsection{Ground-state energy and spin gap}

\begin{figure}
\centerline{\psfig{bbllx=80pt,bblly=250pt,bburx=510pt,bbury=560pt,%
figure=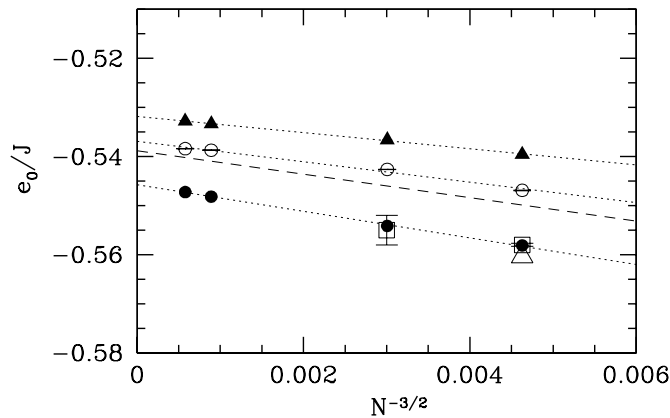,width=80mm,angle=0}}
\caption{\baselineskip .185in \label{fig.taf6}
Ground-state energy per site $e_0 = E_0/N$, in unit of
$J$, as a function of the system size, obtained with
VMC (full triangles), FN (empty dots) and GFMCSR with $p = 7$ (full dots) techniques.
Spin-wave size scaling is assumed and short-dashed lines are
linear fits against $1/N^{3/2}$. The  long-dashed line is the linear spin-wave prediction,
the empty triangle is the $N = 36$ ED result and the empty squares are data
taken from Ref.~\protect\cite{sindzingre}.
}
\end{figure}

In presence of N\'eel long-range order, being the magnon
dispersion relation linear in the wavevector ${\bf k}$, 
the leading finite-size correction to the ground-state energy 
per site is ${\cal O}(N^{-3/2})$:\cite{bernu} this is clearly
shown by the behavior of the finite-size spin-wave results 
in Fig.~\ref{fig.taf6}.
In the same figure the size scaling of the estimates of the ground-state energy
per site obtained with the VMC, the FN and the GFMCSR ($p=7$) techniques is
also reported. The predicted size scaling, fulfilled of course
by the variational wavefunction (\ref{eq.wfhuse}), is also preserved within the
FN and the more accurate GFMCSR technique, thus providing a first clue
on the existence of long-range N\'eel order in the thermodynamic ground
state of the model.
The quality of our results is similar
to the variational one  obtained by Sindzingre and co-workers,\cite{sindzingre}
using a  long-range ordered RVB wavefunction (Sec.~2.4).   
The latter approach
is  almost exact for small lattices, but  the sign problem  is already
present at the variational level, and the calculation  has not been
extended to high statistical accuracy or to $N > 48$.
Our best estimate is  that in the thermodynamic limit 
the ground-state energy per site is
$e_0=-0.5458 \pm 0.0001$ in unit of the exchange coupling.

In the isotropic triangular antiferromagnet, the gap to the first
spin excitation is rather small. Furthermore,
for the particular choice of the guiding wavefunction (\ref{eq.wfhuse}),
the translational symmetry of the Hamiltonian is preserved only if
projected onto subspaces with total $S^{z}$ multiple of three.
Then, we  have studied the gap to the spin $S=3$ excitation
as a function of the system size.
Technically, within our numerical framework, such a spin gap
can be evaluated by performing two simulations in the $S^z=0$ and
$S^z=3$ subspaces. This can be easily done by restricting
the sampling to the configurations $|x\rangle$ in Eq.~(\ref{eq.wfhuse}) with
the desired value of $S^z$. In this case the potential $v(r)$ used
was the same in both subspaces and the variational parameter $\eta$
was found by optimizing the energy in the $S^z=0$ subspace.
 
\begin{figure}
\centerline{\psfig{bbllx=80pt,bblly=250pt,bburx=510pt,bbury=560pt,%
figure=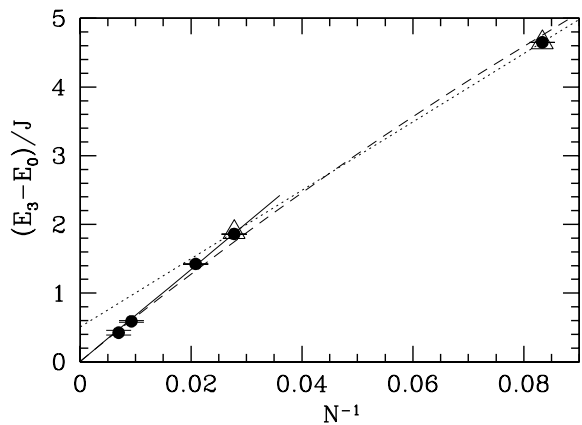,width=80mm,angle=0}}
\caption {\baselineskip .185in \label{fig.taf7}
Size scaling of the spin gap to the $S=3$ excitation obtained with the GFMCSR
($p=7$) technique (full dots). Empty triangles are the ED results,
the long-dashed line is the linear spin-wave prediction,
the dotted line is the linear extrapolation of the $N=12,36$
exact results and
the solid line is the least-squares fit of the GFMCSR data for $N \geq 36$.
}
\end{figure}

As it is shown in Fig.~\ref{fig.taf7}, for the lattice sizes for
which a comparison with ED data is possible, the spin gap estimated with the
GFMCSR technique is nearly exact.
The importance of extending the numerical investigation to clusters
large enough to allow
a more reliable  extrapolation is particularly evident in the same
figure, in which the $N=12$ and 36 exact data extrapolate linearly to a large
finite value. This behavior,
is certainly a finite-size effect and it is corrected by the GFMCSR data
for $N\ge48$, suggesting, strongly, a  gapless excitation spectrum
[$(E_{3}-E_{0})/J=0.002 \pm 0.01$].

\subsubsection{Staggered magnetization}

The GFMC allows us to obtain a very high statistical accuracy on
the ground-state energy, but does not allow us to compute directly
ground-state expectation values $\langle\psi_0 | \hat{O} | \psi_0\rangle$.
A straightforward way to calculate such expectation values is to use the Hellmann-Feynman theorem.
In fact, if the Hamiltonian is perturbed with a term $-\lambda \hat{O}$ the first order correction to the
ground-state energy is, by standard perturbation theory, 
\begin{equation}
E(\lambda)=E_0-\lambda \langle \psi_0|\hat{O}|\psi_0\rangle~.
\label{eq.fopt}
\end{equation}
As a consequence it is possible to evaluate the ground-state expectation value
$\langle\psi_0 | \hat{O} | \psi_0\rangle =- d E(\lambda) / d \lambda |_{\lambda=0}$ estimating the limit
\begin{equation}
\langle\psi_0 | \hat{O} | \psi_0\rangle = - \lim_{\lambda \to 0} \frac{E(\lambda)-E_0}{\lambda}
\end{equation} 
with few computations at different {\em small}  $\lambda$'s.

A further complication for nonexact calculations like the FN or
GFMCSR, is that if the off-diagonal matrix elements 
$\langle x^\prime|\hat{O}|x\rangle$  of the
operator  $\hat{O}$  (in the chosen basis) have signs which are opposite
to those of the product $\psi_G(x^\prime) \psi_G(x)$, they cannot be handled exactly
within the FN because the addition of such a perturbation to the Hamiltonian changes the
nodal surface of the guiding wavefunction.
In that case, in fact, the effective FN Hamiltonian 
associated to the unperturbed Hamiltonian is also affected by the presence of the field and this
leads naturally to the breakdown of Eq.~(\ref{eq.fopt}). 
A way out of this difficulty if to split the operator
$\hat{O}$  into three contributions:
$\hat{O}= \hat{D}+ \hat{O}^{+}+ \hat{O}^{-}$, where $\hat{O}^+$ ($\hat{O}^-$)
is the operator with the same off-diagonal matrix
elements  of $\hat{O}$ when they have
the same (opposite) signs of $\psi_G(x^\prime) \psi_G(x)$,
and zero otherwise,
whereas $\hat{D}$ is the diagonal part of $\hat{O}$.
Then,  we can add to the Hamiltonian a contribution that does not change the
nodes: $\calhat{H}(\lambda) =\calhat{H} -  \lambda (\hat{D}+ 2\,\hat{O}^+) $
for $\lambda > 0$ and
$\calhat{H}(\lambda) =\calhat{H}-\lambda (\hat{D} +2\,\hat{O}^-)$ for $\lambda <0$.
Then the expectation value of the operator $\hat{O}$ can be written as
\begin{equation}
\langle\psi_0 | \hat{O} | \psi_0\rangle = \lim\limits_{\lambda \to 0}
\frac{E(-\lambda)-E(\lambda)}{2 \lambda}~. 
\end{equation}

\begin{figure}
\centerline{\psfig{bbllx=80pt,bblly=195pt,bburx=500pt,bbury=610pt,%
figure=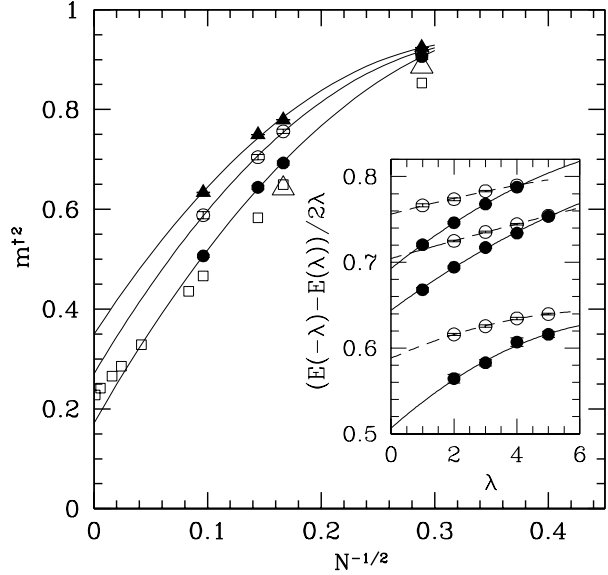,width=80mm,angle=0}}
\caption {\baselineskip .185in \label{fig.taf8}
Size scaling of the order parameter: VMC (full triangles), FN (empty dots),
GFMCSR (full dots), exact data (empty triangles) and finite-size
linear spin wave theory\protect\cite{triangsw} (empty squares).
The inset displays the $\lambda \to 0$
extrapolation for $N>12$.  Lines are quadratic fits in all the plots.
}
\end{figure}

With this method, using the FN and GFMCSR techniques, 
we have calculated the order parameter
\begin{equation}
m^{\dagger 2}= 36 \frac{{\cal M}^2}{N(N+6)}~, 
\end{equation}
where ${\cal M}^2$ is the sublattice magnetization squared.\cite{bernu}
Our results are plotted in Fig.~\ref{fig.taf8}.
For the order parameter the inclusion of many short-range correlations in the SR is
not very important. Then, in order to minimize the numerical effort,
we have chosen to put in the SR conditions the first four correlation functions shown
in Fig.~\ref{fig.taf6}, the order parameter itself and $\hat{\bf S}^{2}$.
While the FN data extrapolate to a value
not much lower than the variational result, the GFMCSR
calculation provides a much more reliable estimate of the order parameter
with no apparent loss of accuracy with increasing sizes.
In this way we obtain for $\hat{m}^{\dagger}$ a value
well below the linear and the second order (which has actually
a {\em positive} correction\cite{miyake}) spin-wave predictions.
Our best estimate is  that in the thermodynamic limit the order parameter
$m^{\dagger}= 0.41 \pm 0.02 $ is reduced by about 59\% from
its classical value. 
This is partially in agreement with the conclusions of the finite-temperature
calculations\cite{elstner} suggesting a ground state with a small but nonzero long-range 
antiferromagnetic order and with series expansions\cite{singh1}
indicating the triangular antiferromagnet
to be likely ordered but close to a critical point.
In our simulation, however, which to our knowledge
represents a first attempt to perform a systematic size-scaling
analysis of the order parameter, the value of $\hat{m}^{\dagger}$ remains
sizable and finite, consistent with a gapless spectrum.

\section{The $J_1-J_2$ Heisenberg Model}
\label{chapj1j2}

In the triangular Heisenberg antiferromagnet, 
frustration is induced by the geometry of the lattice. 
The other possible origin of the frustration comes from competing interactions as in the
so-called $J_{1}{-}J_{2}$ Heisenberg model on the square lattice
defined in Eq.~(\ref{eq.j1j2ham}) of the Introduction.

In the last few years several studies -- including exact diagonalizations of small 
clusters,\cite{schultz,dagotto2} spin-wave\cite{zhong,chandra,trumper} and Schwinger-boson\cite{trumper2}
calculations, series\cite{gelfand,singh2,zith} and large-$N$\cite{read} expansions --
have provided some evidence for the absence of N\'eel order in the ground state 
of the spin-1/2 $J_{1}{-}J_{2}$ Heisenberg model for $0.38 < J_2/J_1 < 0.6$.
However, due to the difficulties related to the
sign problem, a systematic size-scaling of the spin gap has been lacking for many years,
and no definite conclusion on the nature of the non-magnetic phase has been drawn
yet. In particular, an open question is whether the  ground state in the quantum disordered
phase is a RVB spin liquid with no broken symmetry,\cite{figueirido} or 
if it breaks some {\em crystal} symmetries by dimerizing in some special pattern (Sec.~2.4).

In this section, we will address the problem of the quantum phase-transition
to a non-magnetic ground state driven by frustration in the spin-1/2 $J_1{-}J_2$
Heisenberg model by means of the finite-size spin-wave theory, ED
and GFMC techniques. 

\subsection{Finite-size spin-wave results}

The finite-size spin-wave theory for the spin-s $J_1{-}J_2$
Heisenberg model can be derived along the same lines followed for 
the triangular Heisenberg antiferromagnet in Sec.~3.1.
In this section we will only show the main results for
$J_2/J_1<0.5$, assuming  the two sublattice classical N\'eel order 
[Fig.~\ref{fig.neelj1j2}~(a)] in the $xy$-plane.

Applying the unitary transformation (\ref{eq.rotpi}) which rotates 
the spin quantization axis by a angle $\pi$  about the $z$-axis on one of the two
sublattices, setting the order parameter along the $x$-axis, and using the Holstein-Primakoff 
transformation for spin operators at the leading order in $1/s$,
the Fourier transformed spin-wave Hamiltonian results as in Eq.~(\ref{eq.HSW})
with $E_{cl}=-zJs^2N(1-\beta)/2$, $z=4$, $\beta=J_2/J_1$, 
\begin{equation}
A_{\bf k}=1+\beta(\delta_{\bf k}-1)~,\:\: B_{\bf k}=-\gamma_{\bf k}, 
\end{equation}
$\gamma_{\bf k} {=}(\cos\,k_x\,+\,\cos\,k_y)/2$, and $\delta_{\bf k}=
\cos\,k_x\,\cos\,k_y$. 

Similarly to the triangular lattice case, the singular Goldstone modes
[${\bf k}={\bf 0}$ and ${\bf k}=(\pi,\pi)$] cannot be diagonalized by means of
the Bogoliubov transformation but can be recombined to give
the total spin squared $\hat{{\bf S}}^2$ at the leading order in $1/s$:
\begin{equation}
\calhat{H}_{SM}=-J_1szA_{\bf 0}
+J_1z\frac{A_{\bf 0}}{N}\left[(S^x)^2+(S^y)^2+(S^z)^2\right].
\end{equation}
As seen in Sec.~3.1 this term, being $A_0$ positive definite,
favors a singlet ground state and implies the Lieb-Mattis 
property, which has been demonstrated 
only for bipartite Hamiltonians, but nonetheless can be
verified numerically on finite sizes for the $J_1{-}J_2$
Heisenberg model.

As in the triangular case, the above analysis
allows one to derive a variational wavefunction which is both accurate
and easily computable in a quantum Monte Carlo algorithm.
Here we will not repeat the derivation, which follows very closely
the one for the triangular antiferromagnet, and leads to  
the following result for $s=1/2$:
\begin{equation}
\label{eq.gwfj1j2}
|\psi_G\rangle=\sum_{x} S_M(x) \exp{\Big[ \frac{\eta}{2} \sum_{i,j}
v(i-j)S_{i}^zS_{j}^z\Big]}|x\rangle~,
\end{equation}
where $v(r)=1/N\sum_{{\bf q}\neq 0} e^{-i {\bf k}\cdot{\bf r}} g_{\bf k}$
with 
\begin{equation}
\label{eq.gammak}
g_{\bf k}= \frac{v_{\bf k}}{v_{\bf k}-u_{\bf k}}
=1-\sqrt{\frac{1-\beta(1-\delta_{\bf k})+\gamma_{\bf k}} {1-\beta(1-\delta_{\bf k})-\gamma_{\bf k}}}~;
\end{equation}
$|x\rangle$ is an Ising spin configuration specified by assigning
the value of $S_i^z$ for each site and $S_M(x)=(-1)^{N_\uparrow(x)}$ is
the {\em Marshall sign} (Sec.~2.1), 
depending on the number
$N_\uparrow(x)$ of spin up on one of the two sublattices.
The summation is restricted only to the  Ising configurations
with $\sum_{i}S^z_i=0$ in order to enforce the projection onto the $S^z=0$ subspace.
In the following we will use the latter variational wavefunction
as the starting point for more refined quantum Monte Carlo calculations. In particular
for $J_2/J_1=0.5$ we have chosen to work with $\beta=0.4$ in Eq.~(\ref{eq.gammak}).
The possibility to restrict to any total spin projection $S^z=\sum_i S^z_i$ allows one to
evaluate the spin gap by performing two simulations for $S^z=0$ and $S^z=1$.
As in the triangular case, the potential $v(r)$ used was the same in both subspaces
and the variational parameter $\eta$ was found by optimizing the energy
in the $S^z=0$ subspace.
The latter spin-wave variational wavefunction (see Tab.~\ref{tab.erg})
provides a rather good estimate of the ground-state energy for $J_2/J_1 \leq 0.3$. 
Instead such accuracy abruptly decreases instead in the regime of strong frustration, suggesting
a change in the nature of the ground state.

\begin{table}
\tcaption{
Estimates of the ground-state energy per site and
their relative accuracy (in brackets) for $N=36$ and various values of the
$J_2/J_1$ ratio. VMC: variational Monte Carlo. VMCLS: variational Monte Carlo
with LS wavefunction. SRLS: GFMCSR with LS wavefunction and $p=8$ (see text).}
\centerline{\footnotesize\smalllineskip
\begin{tabular}{c c c c c}\\
\hline
$J_2/J_1$  &   VMC        &  VMCLS         &   SRLS          & Exact  \\
\hline
0.0        &-0.6695~(1.4\%)& -0.6756~(0.5\%) &  -0.6789~(0.00\%) & -0.67887\\
0.1        &-0.6284~(1.5\%)& -0.6349~(0.5\%) &  -0.6379~(0.03\%)& -0.63810\\
0.2        &-0.5884~(1.8\%)& -0.5952~(0.6\%) &  -0.5988~(0.04\%)& -0.59905\\
0.3        &-0.5495~(2.3\%)& -0.5574~(0.9\%) &  -0.5619~(0.10\%) & -0.56246\\
0.4        &-0.5120~(3.3\%)& -0.5237~(1.1\%) &  -0.5289~(0.16\%) & -0.52974\\
0.5        &-0.4783~(5.1\%)& -0.4916~(2.4\%) &  -0.5022~(0.32\%) & -0.50381\\
\hline\\
\end{tabular}}
\label{tab.erg}
\end{table}

Within the finite-size spin-wave theory,
we can also gain information about the low-lying excited states. As shown in 
Sec.~3.1.2, in order to stabilize
a low-energy total spin excitation $S$ a magnetic field $h$ in the
$z$-direction must be added to the spin Hamiltonian.
Keeping into account that, for small fields, the classical minimum 
energy configuration is the N\'eel order canted by
an angle $\theta$ along the direction of $h$ (with $\sin\theta=h/2J_1z$),
the finite-size spin-wave expansion is straightforward and leads to
a linearized Hamiltonian as in Eq.~(\ref{eq.HSW})
with the following field-dependent coefficients 
\begin{equation}
A^h_{\bf k}=1+\beta(\delta_{\bf k}-1)+\gamma_{\bf k}\Big(\frac{h}{2J_1z}\Big)^2~,\:\:
B^h_{\bf k}=-\gamma_{\bf k} \Big[1-\Big(\frac{h}{2J_1z}\Big)^2\Big],
\end{equation}
and with a singular part given by
\begin{equation}
{\cal H}_{SM}=-\frac{J_1sz}{2}\;A^h_{{\bf 0}} 
+ \frac{J_1z}{N}  (S^z-Ns\; \sin\theta)^2,
\end{equation}
favoring  a value of $S^{z}$ (in the original spin representation)
consistent  with the applied field, at the classical level. 

The total spin $S=N \langle S^z_{i} \rangle$ of the
excitation can be related to the magnetic field $h$ by means of
the Hellmann-Feynman theorem
\begin{equation}
\langle S^z_{i} \rangle=-\frac{1}{Ns}\frac{\partial}{\partial h}E(h)
= s\frac{h}{2J_1z} \left[1+\frac{1}{2Ns} \sum_{{\bf k}\neq0} \gamma_{\bf k}
\sqrt{\frac{A^h_{\bf k}+B^h_{\bf k}} {A^h_{\bf k}-B^h_{\bf k}}}\right]
\end{equation}
where 
\begin{equation}
\label{eq.EHj1j2}
E(h)=E_{cl}-(sh)^2\frac{N}{4J_1z}-\frac{J_1sz}{2}\left[N(1-\beta)-
\sum_{\bf k}\epsilon^h_{\bf k}\right]
\end{equation}
and $\epsilon_{\bf k}^h=\sqrt{(A^h_{\bf k})^2-(B^h_{\bf k})^2}$.
The final step in order to evaluate
the energy spectrum $E(S)$ is to perform
a Legendre transformation $E(S)=E(h)+hsS$.

Finally, the  spin-wave uniform susceptibility,
$$\chi_{\rm SW}=-1/N \,{\partial^2 E(h)}/{\partial h^2}|_{h=0}~,$$
is, at the leading order in $1/s$,
\begin{equation}
\label{eq.chisw}
\chi_{\rm SW}/\chi_{cl}=1+\frac{1}{2Ns}\sum_{{\bf k}\ne 0}\gamma_{\bf k}
\sqrt{\frac{A_{\bf k}+B_{\bf k}}{A_{\bf k}-B_{\bf k}}}~.
\end{equation}
where $\chi_{cl}=1/2J_1z$.

\subsection{Transition to a quantum disordered state induced by frustration}

\subsubsection{Spin-wave susceptibilities and low-energy spectra}

For the unfrustrated Heisenberg model, even  for $s=1/2$, the spin-wave predictions
are very accurate as far as the energy, the order parameter and the spin uniform susceptibility 
are concerned.\cite{zhong,lavalle}
Turning on $J_2$ the model is increasingly frustrated and one can expect
the spin-wave theory to remain accurate only in the region where
the nature of the order parameter is the same as in the classical
case ($S\to\infty$). Within this analytical approach we can therefore 
detect a non-magnetic phase by looking for the breakdown of
the spin-wave expansion.

\begin{figure}
\centerline{\psfig{bbllx=30pt,bblly=250pt,bburx=525pt,bbury=580pt,%
figure=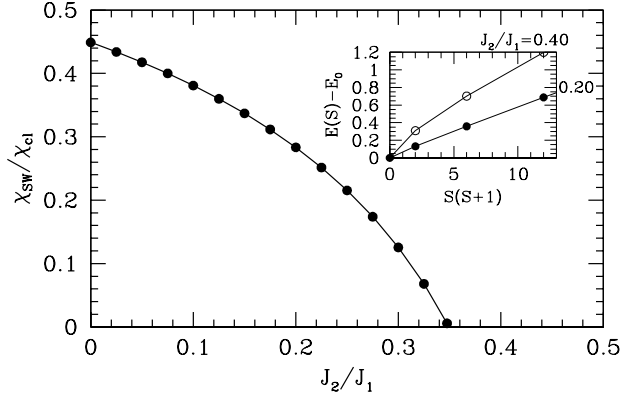,width=80mm,angle=0}}
\caption{\baselineskip .185in \label{fig.chisw}
Spin-wave uniform susceptibility as a function of  $J_2/J_1$ for $s=1/2$. The inset displays
the low-energy spin-wave spectrum for $N=144$ for two values of the $J_2/J_1$
ratio as a function of $|{\bf S}|^2 = S(S+1)$.
}
\end{figure}

As pointed out by Zhong and Sorella,\cite{zhong}
for moderate frustration ($J_2/J_1 <0.2$) 
the linear spin-wave predictions on finite sizes are quite accurate 
for both the energy and the antiferromagnetic order parameter.
Moreover, in this regime, the second order correction\cite{zhong}
leads to an almost exact result.
For large $J_2/J_1$, instead, the second order term does not improve the first order estimate
and a possible breakdown of the spin-wave expansion may occur even well below
the classical transition point $J_2/J_1=0.5$. In particular at the leading order in $1/s$, and
for $s=1/2$ the order parameter vanishes at a critical 
value $(J_2/J_1)_c \simeq 0.38$.\cite{zhong,chandra}

\begin{figure}
\centerline{\psfig{bbllx=20pt,bblly=240pt,bburx=550pt,bbury=550pt,%
figure=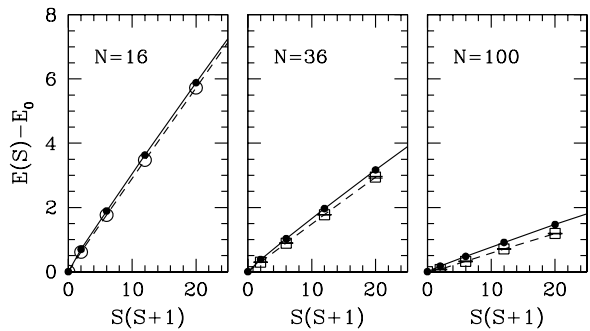,width=80mm,angle=0}}
\caption{\baselineskip .185in \label{fig.tower}
Low-energy excitation spectra as a function
of $|{\bf S}|^2 = S(S+1)$ for $J_2/J_1=0.2$, $s=1/2$ and $N=16, 36, 100$:
spin-wave (full circles and continuous line), exact (empty circles and dashed lines),
GFMCSR with $p=8$  (empty squares and dashed lines).
}
\end{figure}

Analogously, a breakdown of the spin-wave expansion
can be evidenced from the vanishing of the uniform
susceptibility, which is always finite when there is
long-range N\'eel order in the thermodynamic limit 
and vanishes instead in presence  of a finite triplet gap (see Sec.~2.2).
As it is shown in Fig.~\ref{fig.chisw},
the classical uniform susceptibility is strongly renormalized
by the quantum fluctuations for $s=1/2$ at the spin-wave level (\ref{eq.chisw}).
As expected, increasing the frustration such reduction is enhanced
and leads eventually to the vanishing of the susceptibility for $J_2/J_1\simeq 0.35$.
The structure of the finite-size spin-wave excitation spectrum
below and above this critical point is very different (see the inset of
Fig.~\ref{fig.chisw}) with an evident breakdown of the quantum top law (\ref{eq.rotator}),
as well as of the spin-wave approximation scheme, in the non-magnetic phase.
Below the critical point, instead, the spin-wave theory reproduces remarkably well
the exact and 
GFMCSR results for the low-energy part of the spectrum
in the whole range of sizes (see Fig.~\ref{fig.tower}).
Furthermore, as already observed for the triangular antiferromagnet,
increasing the size, the slope of $E(S)$ vs $S(S+1)$ decreases and
gives rise to the collapse of a macroscopic  number of
states with different $S$ on the ground state as
$N\rightarrow\infty$: i.e., a ground state with a broken SU(2)
symmetry.

\subsubsection{Size-scaling of the spin gap}
\label{ssec.sssg}

\begin{figure}
\centerline{\psfig{bbllx=70pt,bblly=250pt,bburx=500pt,bbury=560pt,%
figure=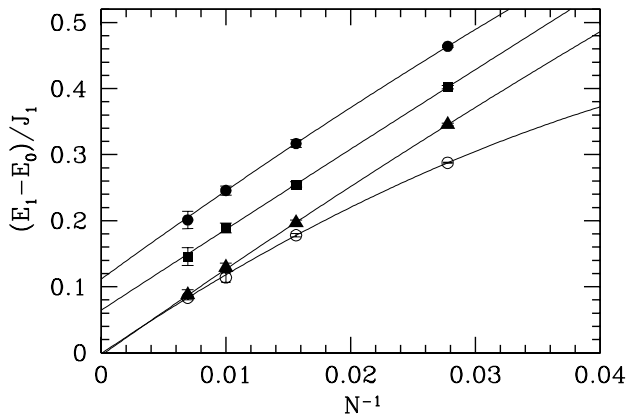,width=80mm,angle=0}}
\caption{\baselineskip .185in \label{fig.gap}
Size scaling of the energy gap to the first $S=1$ spin excitation
obtained with the GFMCSR technique for $J_{2}/J_{1}=0.38$ (full triangles),
0.45 (full squares) and 0.50 (full circles). Data for the unfrustrated ($J_{2}=0
$)
Heisenberg model taken from Ref.~\protect\cite{calandra}, are also shown for
comparison
(empty circles). Lines are weighted quadratic fits of the data.
}
\end{figure}

The spin-wave prediction for the occurrence of a non-magnetic region in the
phase diagram of the $J_1{-}J_2$ Heisenberg model is confirmed by 
our results for the spin triplet gap obtained using the
GFMCSR.
The latter calculation, which extends the recent one by Sorella,\cite{ss} 
has been performed using (\ref{eq.gwfj1j2}) as guiding wavefunction
and including in the SR conditions the energy, 
all $\hat{S}^{z}_{i} \hat{S}^{z}_{j}$ independent by symmetry,
and the antiferromagnetic order parameter. 
The latter, though not improving the accuracy of the
calculation, allows a very stable and reliable simulation
for large $p$.
The new results, extended up to $N=144$,
confirm the previous findings\cite{ss} of a finite spin gap
in the thermodynamic limit 
for $J_{2}/J_{1} > 0.40$ (Fig.~\ref{fig.gap}).
Remarkably,
these results are not an artifact of the chosen guiding wavefunction:
in fact, unlike the FN approximation, the GFMCSR is able to detect
a finite gap in the thermodynamic limit by starting from a
spin-wave like wavefunction (\ref{eq.gwfj1j2}) which is N\'eel ordered and therefore gapless.
This behavior is very different
from the one observed in the case of the $s=1/2$ Heisenberg antiferromagnet on the
triangular lattice (see Fig.~\ref{fig.taf7}) 
where, with the same numerical scheme, a similar guiding wavefunction  and a comparable accuracy
we obtained a gapless excitation spectrum.
Therefore the existence of a gapped phase in the regime of strong frustration
is likely to be a genuine feature of the $J_1{-}J_2$ Heisenberg model.

\subsection{The nature of the non-magnetic phase}

\begin{figure}
\centerline{\psfig{bbllx=150pt,bblly=280pt,bburx=480pt,bbury=455pt,%
figure=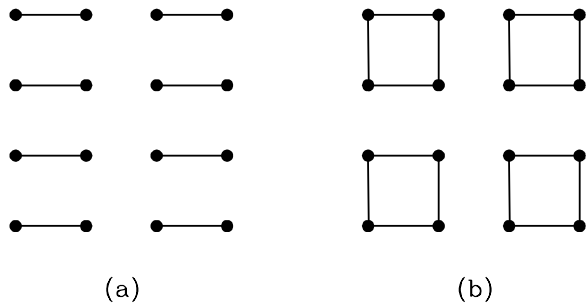,width=80mm,angle=0}}
\caption{\baselineskip .185in
Columnar (a) and plaquette (b) RVB states.}
\label{fig.colplaq}
\end{figure}

In principle either a RVB crystal, with some broken spatial symmetry,
or a homogeneous spin liquid is compatible with a triplet gap in the
excitation spectrum. 
Among the dimerized phases proposed in the literature, the so-called
{\em columnar} and {\em plaquette} RVB
are the states which are the most likely candidates. 
These kind of states can be thought of as a collection of valence bond states
$|\begin{picture}(25,6)(-3,-3)
\put(0,0) {\circle*{3}}
\put(0,0){\line(1,0){18}}
\put(18,0){\circle*{3}}
\end{picture}\rangle =
|\hspace{-3pt}\uparrow\downarrow\rangle{-}
|\hspace{-3pt}\downarrow\uparrow\rangle~$ between neighboring sites arranged
in the patterns shown in Fig.~\ref{fig.colplaq}, where the plaquettes in (b) are the
following rotationally invariant superpositions:
$$
{\Big |}\hspace{-5pt}\begin{array}{cc}
\begin{picture}(10,30)(-5,-5)
\put(0,0) {\circle*{3}}
\put(0,0){\line(0,1){18}}
\put(0,0){\line(1,0){18}}
\put(0,18){\circle*{3}}
\put(0,18){\line(1,0){18}}
\end{picture}
&
\begin{picture}(10,30)(-5,-5)
\put(-1,0) {\circle*{3}}
\put(-1,0){\line(0,1){18}}
\put(-1,18){\circle*{3}}
\end{picture}
\end{array} \hspace{-2pt}{\Big \rangle}
= {\Big |}\hspace{-5pt} \begin{array}{cc}
\begin{picture}(10,30)(-5,-5)
\put(0,0) {\circle*{3}}
\put(0,0){\line(1,0){18}}
\put(0,18){\circle*{3}}
\put(0,18){\line(1,0){18}}
\end{picture}
&
\begin{picture}(10,30)(-5,-5)
\put(-1,0) {\circle*{3}}
\put(-1.1,18){\circle*{3}}
\end{picture}
\end{array}\hspace{-2pt}{\Big \rangle}
+
{\Big |}\hspace{-5pt} \begin{array}{cc}
\begin{picture}(10,30)(-5,-5)
\put(0,0) {\circle*{3}}
\put(0,0){\line(0,1){18}}
\put(0,18){\circle*{3}}
\end{picture}
&
\begin{picture}(10,30)(-5,-5)
\put(-1,0) {\circle*{3}}
\put(-1,0){\line(0,1){18}}
\put(-1,18){\circle*{3}}
\end{picture}
\end{array}\hspace{-2pt} {\Big \rangle}~.
$$
Both the columnar and the plaquette states
break the translation invariance along the $x$ and $y$ directions, but
only the latter preserves the symmetry of interchange
of the two axes.

Read and Sachdev\cite{read} with a field-theoretic large-$N$ expansion,
were the first to conjecture that quantum fluctuations and a next-nearest-neighbor
frustrating interaction could drive the ground state of the square lattice antiferromagnet into a
columnar RVB state and series expansion studies\cite{gelfand,singh2}
have supported over the years this prediction.
Recently, Kotov and co-workers\cite{kotov} with a study that combines
an analytic effective Hamiltonian approach, extended dimer expansions and exact diagonalizations
have presented a body of evidences that has been interpreted as supporting
the columnar scenario. 
Finally, using the GFMCSR with a Density Matrix Renormalization Group guiding wavefunction,
du Croo de Jongh and co-workers\cite{ducroo} have proposed a ground state with
intermediate properties between the plaquette and columnar RVB.

\subsubsection{The method of generalized susceptibilities}

In order to better characterize the
nature of the ground state in the gapped phase, we have checked the occurrence
of some kind of crystalline order,
by calculating the response of the system to operators breaking the
most important lattice symmetries.
As suggested in Refs.~\cite{singh2,zith,santoro}, and also shown in Sec.~2.2, 
the occurence of some kind of crystalline order in the thermodynamic ground state 
can be checked by adding to the Hamiltonian (\ref{eq.j1j2ham})
a term $-\delta \hat{O}$, where $\hat{O}$ is an
operator that breaks some symmetry of $\calhat{H}$.
In fact, if true long-range order  exists
in the thermodynamic ground state,
the finite-size susceptibility
$\chi_{O} = \langle \hat{O} \rangle_{\delta}/N\delta$  has to diverge
with the system size and, in particular, 
it is bounded from below by the system volume 
squared [Eq.~(\ref{eq.chidiv})].
Thus susceptibilities are in principle a very sensitive tool
-- much more than the square of the order parameter --
for detecting the occurrence of long-range order.

Within a numerical technique, the susceptibility
$\chi_{O} = d^2e(\delta)/d\delta^2|_{\delta=0}$ can be calculated with only energy
measurements by computing the ground-state energy per site in presence
of the perturbation for few values of $\delta$ and by estimating numerically
the limit
\begin{equation}
\chi_{O} = \lim_{\delta \to 0} \chi_{O}(\delta)=-\frac{2(e(\delta)-e_0)}{\delta^2}~.
\label{eq.chiex}
\end{equation}

\begin{figure}
\centerline{\psfig{bbllx=60pt,bblly=235pt,bburx=540pt,bbury=560pt,%
figure=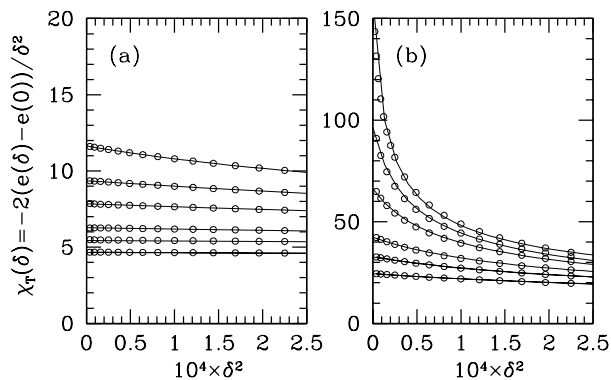,width=80mm,angle=0}}
\caption{\baselineskip .185in \label{fig.chi1d}
Exact results for the $J_{1}{-}J_{2}$ chain:
$\chi_{{\rm T}}(\delta)$ associated to the operator $\hat{O}_{\rm T}$
(breaking the translational invariance)
for $J_2/J_1=0.2$ (a) and $J_2/J_1=0.4$ (b).
Data are shown for $N=12,14,16,20,24,30$
for increasing values of $\chi_{\rm T}(\delta)$. Lines are guides for the eye.
}
\end{figure}

As we have tested in the one dimensional $J_1-J_2$ model,
the numerical study of  long-range order by means of $\chi(\delta)$
is very effective and reliable.
Here a quantum critical point at $J_{2}/J_{1}\simeq 0.2412$
separating a gapless spin-fluid phase from a gapped dimerized ground state
(which is two-fold degenerate and adiabatically connected to
the Majumdar-Ghosh exact solution for $J_{2}/J_{1}=0.5$) 
is rather well accepted.\cite{1dcrit1,1dcrit2,1dcrit3}
As shown in Fig.~\ref{fig.chi1d}, the response
of the system to the perturbation $\delta \hat{O}_{T}$, with
\begin{equation}
\hat{O}_{\rm T}=\sum_{j} e^{i kj}
\hat{{\bf {S}}}_{j} \cdot \hat{{\bf {S}}}_{j+x}~, 
\end{equation}
breaking the translation invariance with momentum $k=\pi$,
is very different below and above the dimer-fluid transition point.
However it is extremely important to perform very accurate calculations at
small $\delta$ to detect the divergence of the susceptibilities for
large system sizes.

\subsubsection{Stability of Plaquette vs Columnar RVB}
\label{ssec.colvsplaq}

\begin{figure}
\centerline{\psfig{bbllx=70pt,bblly=250pt,bburx=500pt,bbury=560pt,%
figure=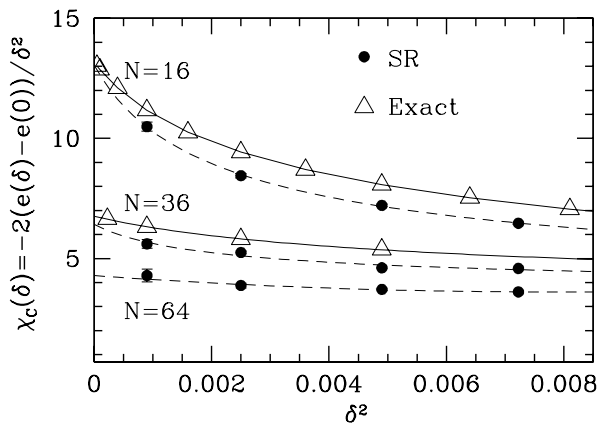,width=80mm,angle=0}}
\caption{\baselineskip .185in \label{fig.chicol}
Exact and GFMCSR calculation of $\chi_{\rm C}(\delta)$ associated
to $\hat{O}_{{\rm C}}$ (columnar dimerization)
for $J_2/J_1=0.5$. Lines are guides for the eye.
}
\end{figure}

As also suggested in a recent paper by Singh and co-workers, \cite{singh2}
the appearance of a columnar state can be probed by using as order parameter
the operator
\begin{equation}
\hat{O}_{\rm C} = \sum_{i} \big( \hat{{\bf {S}}}_{i} \cdot \hat{{\bf {S}}}_{i+x} -
\hat{{\bf {S}}}_{i} \cdot \hat{{\bf {S}}}_{i+y}\big)~,
\label{column}
\end{equation}
where $x=(1,0)$, $y=(0,1)$.
As shown in Fig.~\ref{fig.chicol}, 
the exact diagonalization  results for $N=16$ and $N=36$ indicate that
the susceptibility associated with this kind of
symmetry breaking, $\chi_{\rm C}$, decreases with the system size.
In order to exclude an anomalous size scaling
we have extended the calculation up to $N=64$.
Our quantum Monte Carlo results, which reproduce quite well
the ED data, rule out clearly the columnar dimerization.

The above result is in disagreement with the conclusions of
several series expansion studies.\cite{gelfand,singh2}
However, as stated in Ref.~\cite{singh2},
the series for $\chi_{\rm C}$ are very irregular and do not allow 
a meaningful extrapolation to the exact result.
In our calculation instead, even the ED results for $N \leq 36$,
are already conclusive.

\begin{figure}
\centerline{\psfig{bbllx=33pt,bblly=250pt,bburx=503pt,bbury=560pt,%
figure=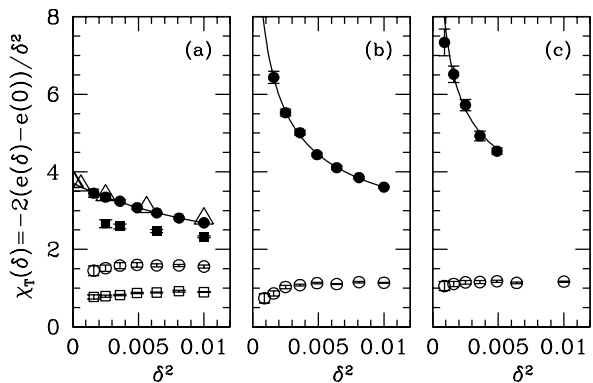,width=80mm,angle=0}}
\caption{\baselineskip .185in \label{fig.comp}
$\chi_{\rm T}(\delta)$ associated to $\hat{O}_{\rm T}$
for $J_2/J_1=0.5$, $N=32$ (a), $N=64$ (b) and $N=100$ (c):
FN (empty squares), GFMCSR (full squares), FN with LS (empty circles),
GFMCSR with LS (full circles), exact (empty triangles).
}
\end{figure}

Having established that the columnar susceptibility is bounded,
it is now important to study the response of the $J_{1}-J_{2}$ model
to a small field coupled to the perturbation
\begin{equation}
\hat{O}_{\rm T}=\sum_{i} e^{i{\bf Q_0}\cdot{\bf r}_i}
\hat{{\bf {S}}}_{i} \cdot \hat{{\bf {S}}}_{i+x}~, 
\end{equation}
with ${\bf Q_0}=(\pi,0)$,
explicitly  breaking the translation invariance of the
Hamiltonian.
The evaluation of $\chi_{{\rm T}}$, with a reasonable accuracy,
is a much more difficult task.
In fact in this case the ED values of the susceptibility
for $N=16$ and $N=32$ increase with the size and much
more effort is then required to distinguish if this behavior
corresponds to a spontaneous symmetry breaking in the thermodynamic limit.
As it is shown in Fig.~\ref{fig.comp}~(a), the FN technique, starting from
a guiding wavefunction without dimer order, is not able to reproduce
the actual response of the system to $\hat{O}_{{\rm T}}$,
even on small sizes.
The GFMCSR technique allows us to get an estimate of the susceptibility
which is a factor of three  more accurate, but not  satisfactory enough.
In order to improve on this estimate,
we have attempted to include in the SR conditions many other,
reasonably simple, correlation functions (such as the spin-spin correlation
functions
$\hat{{\bf {S}}}_{i} \cdot \hat{{\bf {S}}}_{j}$
for $|r_i-r_j|>\sqrt{2}$), but without obtaining a sizable
change  of the estimate of $\chi_{\rm T}$.
In fact, the most effective SR conditions are those obtained with
operators more directly related to the Hamiltonian.\cite{sr,triang}

After many unsuccessful attempts, we have realized that
it is much simpler and straightforward to improve the accuracy
of the guiding wavefunction itself.  
This can be obtained by applying a generalized
Lanczos operator $(1+\alpha\calhat{H})$ to the
variational wavefunction $|\psi_{G}\rangle$,
where $\alpha$ is a variational parameter.
This defines the so-called one Lanczos step (LS) wavefunction.\cite{heeb}
In the present model by using the LS wavefunction,
a clear improvement (by about a factor of 3) 
on the variational estimate of the ground-state energy is obtained
at all strengths of frustration (see Tab.~\ref{tab.erg}).
With this starting point 
the GFMCSR provides an estimate of the ground-state energy
which is basically exact for moderate frustration and remarkable accurate
for $J_2/J_1=0.4$ and 0.5.
More importantly, as shown in Fig.~\ref{fig.comp}~(a), the LS wavefunction allows
a much better estimate of the susceptibility. 
This calculation
was  obtained by including in the SR conditions the energy,
the spin-spin correlation functions up to next-nearest-neighbors,
distinguishing also $\hat{S}^{z}_{i} \hat{S}^{z}_{j}$ and
$(\hat{S}^{x}_{i} \hat{S}^{x}_{j}+\hat{S}^{y}_{i} \hat{S}^{y}_{j})$
($p=4$).
The mixed averages of these correlation functions can be computed over
both the wavefunction $|\psi_{G}\rangle$ and the LS wavefunction
$(1+\alpha\calhat{H})|\psi_{G}\rangle$ during the same
Monte Carlo simulation.
Thus with a LS wavefunction one can also easily double the number of constraints
that are effective to improve the accuracy of the method ($p=8$).
In this case, we have tested that it is irrelevant to add further
long-range correlation functions in the SR conditions even for large size.

By increasing the size [see Figs.~\ref{fig.comp}], 
the response of the system is very strongly enhanced,
in very close analogy with the one dimensional model
in the dimerized phase  [see Fig.~\ref{fig.chi1d}~(b)].
This is obtained only with the GFMCSR technique, since as shown
in Fig.~\ref{fig.comp},
 the combination of FN and Lanczos step alone, is not capable of
detecting these strongly enhanced correlations.
For $N=100$ the GFMCSR increases by more than one order of magnitude
the response of the system to the dimerizing field.
Of course, for a definite conclusion, one should check whether the susceptibility diverges
as the volume squared, as implied by Eq.~(\ref{eq.chidiv}). However,
in order to obtain quantitatively reliable zero-field extrapolations (\ref{eq.chiex}),
the limit of very small fields has to be reached.
This is in general possible within exact diagonalization 
(see Fig.~\ref{fig.chi1d}) but it is rather difficult within
a stochastic technique like the GFMC which is always affected by a statistical
error. 

\section{Conclusions}
\noindent

In this paper we have studied the interplay between frustration and
zero-point quantum fluctuations in the ground state of the triangular 
and $J_1{-}J_2$ Heisenberg antiferromagnets.
These frustrated systems are the simplest examples of two-dimensional
spin models in which quantum effects may be strong enough to destroy
the classical N\'eel order, thus stabilizing a ground state
with symmetries and correlations different from their classical counterparts.
For this reason, in the last few years, they have attracted much theoretical interest even if 
a general consensus on the nature of their ground state has not yet been achieved.
With this work, by using several techniques including
the Green function Monte Carlo with Stochastic Reconfiguration,\cite{ss,sr}
a quantum Monte Carlo method recently developed to keep under control 
the sign problem, we have put on firmer grounds
the conclusions on the ground-state properties of these frustrated models.

Despite the fact that the spin-half Heisenberg antiferromagnet on the triangular lattice was the first
historical candidate for a non-magnetic ground state,\cite{anderson1,fazekas} 
all our results point toward the existence of zero-temperature 
long-range N\'eel order. 
In fact, our quantum Monte Carlo simulations  
provide  robust evidences for a gapless spectrum and for a value of 
the order parameter that, although reduced (by about 59\%) 
with respect to the classical case, remains finite in the thermodynamic limit.
This is partially in agreement with the conclusions of  finite-temperature
calculations\cite{elstner} suggesting a ground state with a small but nonzero 
long-range antiferromagnetic order and with series expansions studies\cite{singh1}
indicating the triangular antiferromagnet to be likely ordered but 
close to a critical point.
However, in our simulation, which to our knowledge
represents the first attempt to perform a systematic finite-size scaling
analysis, the value of the thermodynamic order parameter is sizeable, indicating 
the presence of stable long-range order.
Moreover, the accuracy of the finite-size spin-wave
predictions indicates that the spin-wave theory is a reliable analytical
approximation to describe the ground-state properties of the present model.
In particular, the effectiveness of the spin-wave theory in reproducing on finite sizes the
low-energy excitation spectrum provides further support
to the existence of long-range N\'eel order in the ground state, suggesting also
that the value of the uniform spin susceptibility should be very
close to the spin-wave result.
We believe that our results, together with the clear indications recently provided by 
Bernu and co-workers\cite{bernu} with a symmetry analysis of the low-energy 
excitation spectra, 
finally solve the issue of the ordered nature of the ground state of the Heisenberg 
antiferromagnet on the triangular lattice.

The effects of quantum fluctuations are more remarkable in the $J_{1}{-}J_{2}$
Heisenberg model, where the combined effect of frustration and zero-point motion
interferes with the mechanism of spontaneously broken symmetry, giving rise to
a non-magnetic ground state of purely quantum-mechanical nature.
In fact, our spin-wave, exact diagonalization, and quantum Monte Carlo results
indicate that quantum fluctuations are able to melt the antiferromagnetic 
long-range order in the regime of strong frustration, 
driving the ground state into a quantum disordered phase at $J_2/J_1 \simeq 0.4$.
In addition, with Lanczos and quantum Monte Carlo calculations we have studied 
the susceptibilities for the most important crystal symmetry breaking operators. 
Our results, while casting serious doubt on the conclusions 
of series expansion studies,\cite{singh2,kotov} 
indicate the {\em plaquette} RVB, with spontaneously broken translation 
symmetry and no broken rotation symmetry, as a more plausible ground state 
in the non-magnetic phase. However further investigation is needed to clarify the nature
of the ground state in the non-magnetic phase and a more refined numerical investigation
is now in progress\cite{nature?}.
In the ordered phase, instead, similarly to the triangular case, we find 
a remarkable agreement between
the spin-wave low-energy excitation spectrum and the exact and quantum
Monte Carlo results.
This suggests that the value of the uniform spin susceptibility
should be very close to the spin-wave prediction up to $J_2/J_1\simeq 0.30$.

Our results could be also verified experimentally 
on the novel realizations of these frustrated models, like the triangular K/Si(111):B 
interface,\cite{weitering} and the ${\rm Li}_2{\rm VOSiO}_4$, $ {\rm Li}_2{\rm VOGeO}_4$ 
compounds,\cite{melzi}
quite recently argued to be well described by a spin-half $J_1{-}J_2$ Heisenberg 
model on the square lattice. 
Forthcoming measurements under pressure\cite{carretta}
could also allow one to tune the $J_2/J_1$ ratio and to investigate 
the properties of these systems in various regimes of frustration.

\nonumsection{Acknowledgements}
\noindent

I wish to thank Sandro Sorella for three years of enlightening teachings.
It is also a pleasure to acknowledge the stimulating collaboration with
Federico Becca, Alberto Parola, Giuseppe Santoro, 
and Adolfo Trumper. Thanks to Valerio Tognetti, Attilio Rigamonti
and Pietro Carretta for suggestions and stimulating discussions,
and to Claire Lhuillier for useful advice and for a careful 
reading of the Ph.D. thesis from which the present work has been adapted.
Special thanks to Francesca Ferlaino for discussions, 
support and warm encouragement.

This work was partially supported by MURST (COFIN99).

\nonumsection{References}

\end{document}